\documentclass[twocolumn,prb,floatfix,superscriptaddress,nobibnotes]{revtex4-2}
\usepackage{graphicx}
\usepackage{enumerate}
\usepackage[colorlinks=true,citecolor=blue]{hyperref}
\usepackage{textcomp}
\usepackage{amsmath}
\usepackage{amssymb}
\usepackage{pgf}
\usepackage{subfigure}
\usepackage[english]{babel}
\usepackage[normalem]{ulem}
\usepackage{tabularray}
\usepackage{xcolor}
\usepackage{array}
\usepackage{float}

\def\ket#1{| #1\rangle}
\newcommand{\braket}[2]{\langle #1 | #2 \rangle}

\newcommand{\hc}{{\rm H.c.}}

\graphicspath{{./}{./Figures/}}

\begin{document}
\title{Roadmap towards Majorana qubits and nonabelian physics in  quantum dot-based minimal Kitaev chains}


\author{Athanasios Tsintzis}
\affiliation{Division of Solid State Physics and NanoLund, Lund University, S-22100 Lund, Sweden}

\author{Rubén Seoane Souto}
\affiliation{Division of Solid State Physics and NanoLund, Lund University, S-22100 Lund, Sweden}
\affiliation{Center for Quantum Devices, Niels Bohr Institute, University of Copenhagen,  2100 Copenhagen, Denmark}
\affiliation{Departamento de Física Teórica de la Materia Condensada, Condensed Matter Physics Center (IFIMAC) and Instituto Nicolás Cabrera, Universidad Autónoma de Madrid, 28049 Madrid, Spain}
\affiliation{Instituto de Ciencia de Materiales de Madrid (ICMM), Consejo Superior de Investigaciones Científicas (CSIC),
Sor Juana Inés de la Cruz 3, 28049 Madrid, Spain.}

\author{Karsten Flensberg}
\affiliation{Center for Quantum Devices, Niels Bohr Institute, University of Copenhagen, 2100 Copenhagen, Denmark}
\affiliation{NNF Quantum Computing Programme, Niels Bohr Institute, University of Copenhagen, 2100 Copenhagen, Denmark}

\author{Jeroen Danon}
\affiliation{Department of Physics, Norwegian University of Science and Technology, NO-7491 Trondheim, Norway}

\author{Martin Leijnse}
\affiliation{Division of Solid State Physics and NanoLund, Lund University, S-22100 Lund, Sweden}
\affiliation{Center for Quantum Devices, Niels Bohr Institute, University of Copenhagen, 2100 Copenhagen, Denmark}

\begin{abstract}
The possibility to engineer artificial Kitaev chains in arrays of quantum dots coupled via narrow superconducting regions has emerged as an attractive way to overcome the disorder issues that complicate the realization and detection of topological superconducting phases in other platforms. Although a true topological phase would require long chains, already a two-site chain realized in a double quantum dot can be tuned to points in parameter space where it hosts zero-energy states that seem identical to the Majorana bound states that characterize a topological phase. These states were named ``poor man's Majorana bound states'' (PMMs) because they lack formal topological protection. In this work, we propose a roadmap for next-generation experiments on PMMs. The roadmap starts with experiments to characterize a single pair of PMMs by measuring the Majorana quality, then moves on to initialization and readout of the parity of a PMM pair, which allows measuring quasiparticle poisoning times. The next step is to couple two PMM systems to form a qubit. We discuss measurements of the coherence time of such a qubit, as well as a test of Majorana fusion rules in the same setup. Finally, we propose and analyse three different types of  braiding-like experiments which require more complex device geometries. Our conclusions are supported by calculations based on a realistic model with interacting and spinful quantum dots, as well as by simpler models to gain physical insight. Our calculations show that it is indeed possible to demonstrate nonabelian physics in minimal two-site Kitaev chains despite the lack of a true topological phase. But our findings also reveal that doing so requires some extra care, appropriately modified protocols and awareness of the details of this particular platform. 
\end{abstract}

\maketitle

\section{Introduction}

Majorana bound states (MBSs)~\cite{Wilczek2009, Alicea_RPP2012, Leijnse_Review2012, AguadoReview, BeenakkerReview_20} are zero-energy excitations described by Hermitian operators. They have been predicted to exhibit exciting new physics, such as nonabelian and nonlocal properties, which could potentially be harnessed in topological quantum computation~\cite{NayakReview2008, Sarma2015, Lahtinen2017, Marra2022}. A simple theoretical model hosting MBSs is the Kitaev chain, a tight-binding model of a spinless $p$-wave topological superconductor~\cite{Kitaev_2001}. By now, there have been many theoretical proposals (see Refs.~\cite{Lutchyn_PRL2010, Oreg_PRL2010, PhysRevB.88.020407, Hell2017, Pientka2017, Vaitiekenas2020, flensberg2021engineered} for a few examples) and substantial experimental efforts (see Refs.~\cite{Mourik_science2012, deng2012anomalous, finck2013anomalous, NadjPerge2014, deng2016majorana, Nichele_PRL2017, lutchyn2018majorana, Fornieri2019, Ren2019, Vaitiekenas2020, Aghaee2022, Yazdani2023} for a few examples) aiming at creating and detecting MBSs in various systems for which the Kitaev chain is an adequate low-energy description.  Despite experiments showing  signatures expected for MBSs in tunneling spectroscopy, the possible emergence of nontopological Andreev bound states (ABSs) due to disorder ~\cite{Prada_PRB2012,Kells_PRB12,Liu2012,Liu2017, Moore_PRB18,reeg2018zero,Awoga_PRL2019,Vuik_SciPost19,Pan_PRR20,Prada_review,hess2021local} generally makes the interpretation  challenging. Measurements of nonabelian and nonlocal physics would constitute much stronger evidence of topological MBSs, but still seem beyond experimental reach, with measurements of lifetimes or coherence properties of MBS candidates still lacking (unlike for nontopological ABSs~\cite{Janvier_Science_2015,Hays_PRL_2018,Hays_science_2021,Hays_Nature_2020}). 

One way to remedy the effects of disorder and material imperfections is to create a Kitaev chain with an array of quantum dots (QDs) coupled via superconducting segments~\cite{Sau_NatComm2012, Leijnse_PRB2012, Fulga2013}. Although long chains are needed to reach a true topological phase, already a minimal Kitaev chain with only two QDs separated by a superconductor can host states similar to  MBSs~\cite{Leijnse_PRB2012}. These states were called poor man's MBSs (PMMs) because they only appear in fine-tuned points (sweet spots) of the parameter space and thus lack topological protection. However, PMMs should, in principle, share all of the exotic properties of topological MBSs, including the nonlocal and nonabelian properties.

In the simplest QD-based two-site Kitaev chain model~\cite{Leijnse_PRB2012}, two spin-polarized QDs are coupled via two processes that involve the superconducting segment, namely elastic cotunneling (ECT) and crossed Andreev reflection (CAR). The sweet spot where PMMs appear is found for equal strengths of CAR and ECT and QD orbitals tuned to zero energy (the middle of the superconducting gap). Since CAR is suppressed for parallel QD spins \cite{Recher2001, Hofstetter2009, Herrmann2010, Schindele2012} while ECT is suppressed for anti-parallel spins, Ref.~\cite{Leijnse_PRB2012} suggested controlling the angle between the QD spins to fine-tune the system to the sweet spot. However, controlling the angle between the QD spins is not straightforward in practice. Another problem with this spinless model is that the QDs in a real system will never be completely spin-polarized and the charging energy in the QDs will also play a role. It is thus fair to wonder whether and how PMMs would emerge in a system of partly spin-polarized and interacting QDs.

An alternative approach to control the relative magnitudes of ECT and CAR without the need to fine-tune the angle between the QD spins was suggested in Ref.~\cite{PhysRevLett.129.267701}. If the two QDs couple via an ABS inside the superconductor, quantum interference between different tunneling processes causes both CAR and ECT to depend on the energy of the ABS, but in different ways. Thus, all that is needed to reach the PMM sweet spot is a global magnetic field together with spin--orbit coupling and control of the ABS energy. 

The effects of interactions, finite Zeeman splittings and strong coupling to the ABS were studied in Ref.~\cite{Tsintzis2022}. Under these realistic conditions, although there are still exact degeneracies between ground states with even and odd electron number parity, there are no points in parameter space where these degeneracies are associated with excitations that are identical to true MBSs. This necessitates introducing a Majorana quality measure that ideally could be accessed in experiment. One such measure is the Majorana polarization (MP)~\cite{Aksenov2020, Sedlmayr2015, Sedlmayr2016}. It was shown in Ref.~\cite{Tsintzis2022} that it is indeed possible to realize PMMs that are close to true MBSs, as indicated by a close-to-ideal MP, but the requirements on the parameters for doing so are non-trivial and it is also possible to end up with low-quality PMMs. Furthermore, it is not clear what is actually needed in terms of PMM quality to be able to explore  nonlocal and nonabelian physics.

Recently, a series of pioneering experiments~\cite{Wang2022, Dvir2023, Wang2022a, Bordin2022, Bordin2023} have demonstrated a high degree of control of the ECT and CAR magnitudes, and transport spectroscopy results that seem to be compatible with the analyses of Refs.~\cite{PhysRevLett.129.267701, Tsintzis2022}. In particular, Ref~\cite{Dvir2023} showed both local and nonlocal tunnel spectroscopy data that seem fully consistent with PMMs. Therefore, it appears to us that the time has now come to move on to the next generation of Majorana experiments based on PMMs.

The purpose of this paper is to sketch a roadmap for this next generation of Majorana experiments, which has the demonstration of nonabelian physics as end destination. Before going into the details, we start with a short overview of the different experiments we propose in this roadmap. We first give a brief introduction to the underlying physics and general goal of each experiment, before discussing the special problems and possibilities offered by performing them with PMMs in the QD platform. 

\subsection{Majorana quality assessment}
Most simple measurements, like local spectroscopy, cannot unambiguously distinguish between MBSs and the (non-topological)  ABSs that can appear because of disorder and/or smooth potential variations~\cite{Prada_PRB2012,Kells_PRB12,Liu2012,Liu2017, Moore_PRB18,reeg2018zero,Awoga_PRL2019,Vuik_SciPost19,Pan_PRR20,Prada_review,hess2021local}. In fact, one can argue that in a finite and disordered system, there is no fundamental difference between these two types of states, because a zero-energy ABS can be decomposed into two MBSs. What we call a MBS is then an ABS that decomposes into two spatially separated MBSs, such that a local experimental probe only interacts with one of them. In the ABS limit, on the other hand, the two MBSs are localized in the same region in space and any experimental probe will couple to both of them. Clearly, there is a gray zone in between these two limits where we have partially separated MBSs~\cite{Vuik_SciPost19, Prada_review, Zeng2020, Liu2021, Tian2021, Ricco2022, Marra2022b, Zeng2022}. Whether it is possible to carry out nonlocal and nonabelian operations in this regime will depend on the details and timescales of the experimental protocol. It is therefore desirable to have a measure of Majorana quality. However, it is not clear what the best quality measure is; this will in general depend on for what experiment one wants to asses the Majorana quality. In this work, we will mostly use the MP~\cite{Aksenov2020,Sedlmayr2015,Sedlmayr2016} (Sect.~\ref{sec:DQDMP}) as a local Majorana quality measure, and we will show how it affects protocols aiming to measure nonlocal and nonabelian properties in our PMM setup. 

Compared with local spectroscopy, nonlocal tunnel spectroscopy allows extracting more information about a bound state~\cite{Gramich2017, Rosdahl2018, Menard2020, Danon_non-local_2020, Pan2021, Pikulin2021, Aghaee2022, Dvir2023}. In a slightly more complicated setup where a QD is tunnel coupled to the part of the (possibly) topological system where a zero-energy state is localized, it is possible to estimate the MBS localization~\cite{deng2016majorana, Clarke_PRB2017, deng2018nonlocality}. If the QD level is tuned to zero energy, it interacts with the zero-energy state. If the zero-energy state is a single isolated MBS, the resulting hybridized QD-MBS leaves a single zero-energy state, while coupling to a second MBS (the ABS limit) results in a splitting.  The spectrum can be measured by a single tunnel probe coupled to the QD. In this work, we investigate this type of spectroscopy in the PMM setup in Sect.~\ref{sec:pradaclarke}, which requires coupling a third QD to the double QD hosting the PMMs. 

\subsection{Majorana initialization and readout}
Two MBSs form a single fermionic state. With no other low-energy states or excited quasiparticles, the occupation of this state (empty or full) correlates directly with the parity of the total electron number (even or odd). For a truly topological system with perfectly separated MBSs two things must hold: The even and odd ground states are perfectly degenerate (there is a fermionic state at zero energy) and no local measurement can read out the occupation of that state. Different ways have been proposed to initialize and read out the state by controlled breaking of the topological protection, see, for example, Refs.~\cite{Alicea_NatPhys2011, van_Heck_NJP2012, Aasen_PRX2016, Hell_milestonesPRB16, Flensberg_PRL2011, Plugge_NJP2017, Karzig_PRB2017, munk2020parity, steiner2020readout}. In this work, we will show in Sect.~\ref{sec:initialization} that the PMM's lack of topological protection is not only a bug but also a feature that allows for easy initialization and readout. We will also discuss how to measure the lifetime of such a state, which will be the limiting factor for the relaxation time of a PMM qubit. Such experiments will require coupling one of the QDs in the PMM setup to, for example, a charge detector~\cite{Elzerman2004, Reilly2007, Barthel2009, Volk2019, Liu2021a} or to a circuit measuring the quantum capacitance~\cite{Petersson2010, Lambert2016, Vigneau2023}. 

\subsection{Majorana qubits}
Because the state encoded in a pair of MBSs corresponds to the parity of the electron number which is a conserved quantity, the minimal way to encode a useful qubit is in the two-level subspace spanned by the state of four MBSs with fixed total parity~\cite{Bravyi2006, Leijnse_Review2012, Alicea_RPP2012, Sarma2015}. Arbitrary (unprotected) single-qubit rotations can be performed by introducing an energy associated with the occupation of the fermionic state encoded in different MBS pairs. Such controllable MBS couplings can, for example, be based on direct overlap of MBS wavefunctions~\cite{Alicea_NatPhys2011, Hell2017, Hell2017b} or Coulomb interactions~\cite{van_Heck_NJP2012, Aasen_PRX2016, Hell_milestonesPRB16}. In Sects.~\ref{sec:coupledDQDs} and~\ref{sec:qubits} we will discuss how to realize MBS qubits by coupling two PMM systems, either by a direct coupling between one QD in each system or by a coupling via a superconducting region. We calculate the spectrum as a function of this coupling, show how this crucially depends on the MP, and discuss the effects on single-qubit gates. This setup also allows measurements of both relaxation and coherence times. These measurements will be a benchmark on how well protected these nontopological MBSs actually are, which is crucial information to judge the feasibility of the nonabelian experiments to follow.   

\subsection{Majorana fusion}
The process of bringing together nonabelian anyons to measure their joint quantum state is known as fusion, and the fusion rules~\cite{NayakReview2008} describe the possible outcomes of such a measurement. MBSs are a realization of so-called Ising anyons~\cite{NayakReview2008}, for which the fusion rules state that pairs can either annihilate (measurement outcome gives an empty fermionic state in the language used above) or combine into a regular fermion (measurement outcome gives a filled fermionic state). The simplest nontrivial experimental test of this physics requires four MBSs, where one configuration of pairings is chosen for intialization, then a different pairing is chosen for readout~\cite{Aasen_PRX2016, Hell_milestonesPRB16, Clarke2017Apr, Beenakker2019, Zhou2020, Zhou2022, Souto2022, Liu2022_PMMfusion}. 

The fusion rules encode the possible measurement outcomes. In this work, we briefly discuss the fusion of PMMs in interacting QDs in Sect.~\ref{sec:fusion} in a protocol similar to another recent proposal~\cite{Liu2022_PMMfusion}, requiring the same setup and control capabilities as in the qubit experiments in Sect.~\ref{sec:qubits}. Although fusion of PMMs indeed seems experimentally feasible, it is not sensitive to the Majorana quality~\cite{Clarke2017Apr} and we show that the result is independent of the MP as long as one can finetune to a point where the uncoupled PMM systems exhibit a perfect even--odd degeneracy.

\subsection{Nonabelian operations and braiding}
The true hallmark feature of nonabelian anyons is that the result of particle exchange (braiding) is described by nonabelian representations of the braid group~\cite{NayakReview2008}, such that the order of exchange operations matters. The robustness of the result of a braid operation---for a true topological system it depends only on which particles are exchanged, not on the details of the exchange---is the foundational property of topological quantum computing schemes~\cite{NayakReview2008, Karzig_PRB2017, Sarma2015, Marra2022}. It might be possible to  measure the nonabelian exchange properties of MBSs by actual physical exchange of MBSs~\cite{Ivanov2001, Alicea_NatPhys2011}, and there are recent proposals for doing so in QD-based Kitaev chains~\cite{Boross_2023}. However, most recent proposals in different superconductor--semiconductor hybrid platforms rely instead on sequences of operations that can be proven to be mathematically equivalent to braiding, including  controlled additions and removal of single electrons (which we will refer to as charge-transfer-based nonabelian operations)~\cite{Flensberg_PRL2011, Souto2020, Krojer_PRB2022}, sequences of measurements of MBS pairs (measurement-based braiding)~\cite{Bonderson_PRL2008, Plugge_NJP2017, Karzig_PRB2017}, and cyclic tuning of hybridization between different MBSs (hybridization-induced braiding)~\cite{Clarke2011, van_Heck_NJP2012, Karzig2015, Aasen_PRX2016, Hell_milestonesPRB16, Hell2017, Clarke2017Apr}. 

In Sect.~\ref{sec:braiding} we propose, analyze and numerically test setups and protocols for all these three types of nonabelian protocols, each of which come with different advantages and difficulties. They all build directly on the ingredients introduced in earlier sections, although more complex geometries are needed. Our focus here is on deviations from the ideal MBS braiding results induced by the imperfect MBS quality (signaled by the MP) which is unavoidable in a realistic PMM system. We find that, in contrast to the fusion protocol, although the result of a braiding operation is in general a nonabelian operation, it only comes close to the topological MBS result for close-to-ideal MP. Thus, we conclude that a braiding experiment will be the real test of the similarities between PMMs and true topological MBSs. 

\begin{figure}[b] \centering
\includegraphics[width=1.0\linewidth]{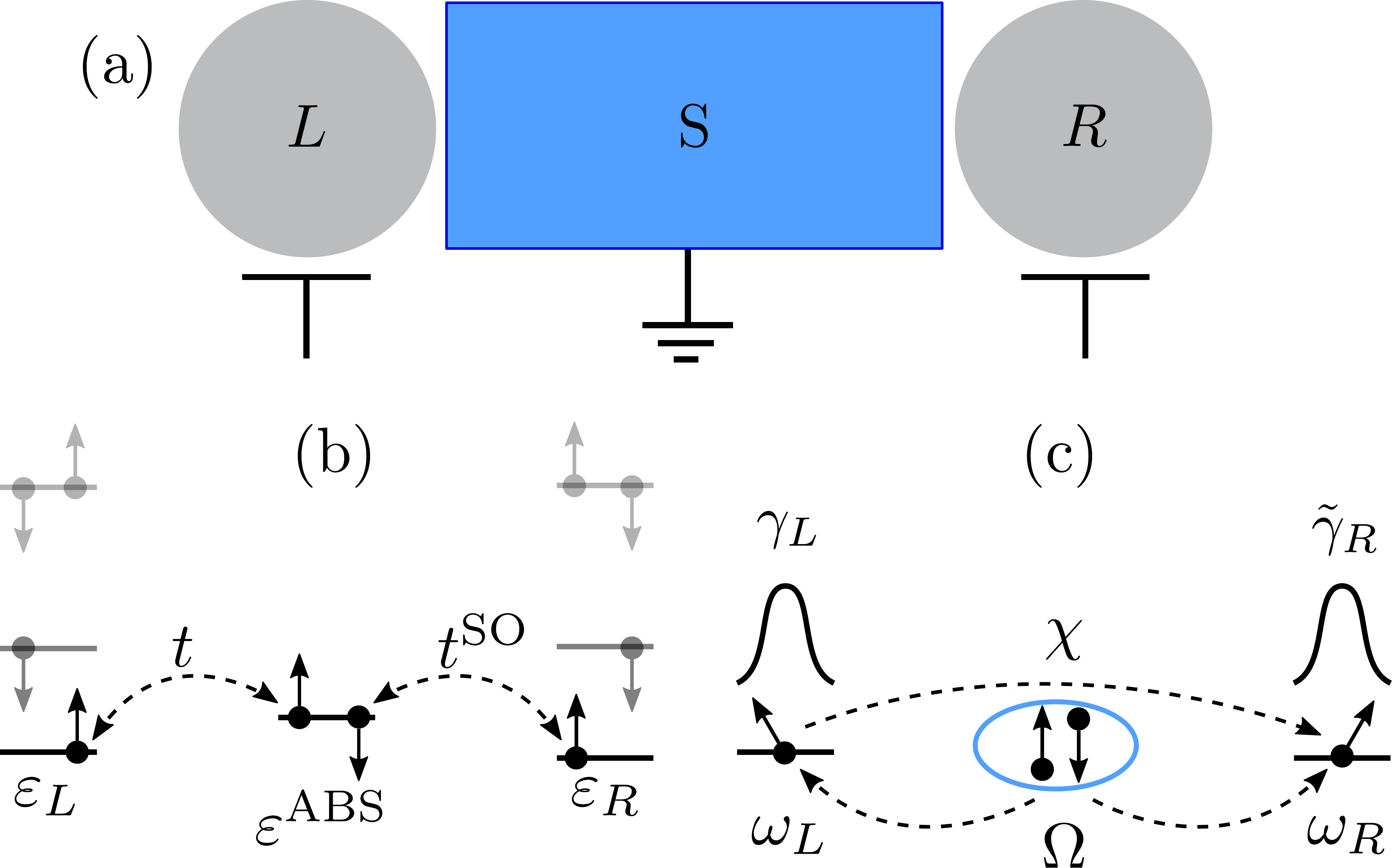}
\caption{(a) Sketch of a PMM system, with two QDs coupled via a grounded superconductor. We will later consider setups where one or both QDs couple to normal metal leads, charge detectors or additional PMM systems. (b) Energy level diagram showing the orbitals of the individual QDs and the ABS, as well as different tunnel processes connecting the QDs and the ABS. (c) Sketch of the spinless model where QDs $L$ and $R$ are coupled via ECT and CAR. The Majorana operators $\gamma_L$ and $\tilde{\gamma}_R$ describe the degenerate ground state when the system is tuned to a sweet spot.
}\label{Fig1}
\end{figure}

\section{Majorana bound states in a double quantum dot}\label{sec:DQD}
Here, we introduce and discuss the basic system with two QDs coupled via a superconductor as sketched in Fig.~\ref{Fig1}(a) and briefly explain how PMMs appear in this system. In later parts of the paper we will add additional ingredients to this basic setup. Throughout the paper we set $e = \hbar = k_B = 1$.

\subsection{Model with interacting spinful quantum dots}\label{sec:modelspin}
We will base most of our conclusions in this work on a model that includes spin and Coulomb interactions on the QDs and allows for strong coupling to the superconductor~\cite{Tsintzis2022}. Figure~\ref{Fig1}(b) shows a sketch of the states of the two QDs and their couplings to the superconductor, which we describe by the Hamiltonian $H=H_\mathrm{QDs} + H_\mathrm{ABS} + H_\mathrm{T}$. Here, the two QDs are described by

\begin{equation}
\label{eq:QDs}
H_\mathrm{QDs} = \sum_{\sigma, j} \varepsilon_{j \sigma} n_{j \sigma} + \sum_j U_j n_{j \uparrow} n_{j \downarrow},
\end{equation}
where $d_{j\sigma}^\dagger$ creates an electron with spin $\sigma=\uparrow, \downarrow$ in QD $j=L,R$ with occupation $n_{j \sigma} = d_{j \sigma}^\dagger d_{j \sigma}$, single-particle orbital energy $\varepsilon_{j \sigma }$ which includes the Zeeman energy $\varepsilon_{j\uparrow/\downarrow} = \varepsilon_j \pm E_{{\rm{Z}}j}/2$, and Coulomb charging energy $U_j$. We assume that the orbital spacing is large enough that we can restrict the model to a single orbital on each QD.

The coupling between the QDs via the superconductor is assumed to be dominated by a discrete ABS. This might be a subgap state in a narrow region of proximitized semiconductor~\cite{PhysRevLett.129.267701, Wang2022, Dvir2023, Wang2022a, Bordin2022} or a third QD that is in turn strongly coupled to a bulk superconductor~\cite{Tsintzis2022}. For definiteness, we choose
\begin{eqnarray}\label{eq:ABS}
H_\mathrm{ABS} = \sum_{\sigma} \varepsilon^\mathrm{ABS}_\sigma n_{\sigma}^\mathrm{ABS} +\Delta  c_{\uparrow}^\dagger c_{\downarrow}^\dagger + \Delta^* c_\downarrow c_\uparrow,
\end{eqnarray}
where $c_{\sigma}^\dagger$ is the electron creation operator in the superconductor, $n_{\sigma}^\mathrm{ABS} = c_{\sigma}^\dagger c_{\sigma}$ and   $\varepsilon^\mathrm{ABS}_{\uparrow/\downarrow} = \varepsilon^\mathrm{ABS} \pm E_{{\rm{Z}}}^\mathrm{ABS}/2$.
We assume that due to the proximity of the superconductor on-site interactions are strongly suppressed for the ABS. We for now assume $\Delta$ to be real, but will discuss the general case later when considering setups with more than one superconductor.

The coupling between the QDs and the ABS is described by
\begin{align}
\begin{split}
H_\mathrm{T} {} & {} = 
 \sum_{\sigma} s_\sigma \Big[ t^\mathrm{SO}_{L} d_{L \sigma}^\dagger c_{\bar{\sigma}} + t^\mathrm{SO}_{R}   c_{\sigma}^\dagger d_{R \bar{\sigma}} \Big] \\
{} & {} + \sum_{\sigma, j} t_{j} d_{j \sigma}^\dagger c_{\sigma}  +  \hc
\end{split}
\label{eq:HT}
\end{align}
Here, $s_{\uparrow, \downarrow} = \pm 1$, $t_{j}$ is the amplitude for spin-conserving tunneling between QD $j$ and the ABS and $t_j^{{\rm SO}}$ is the  amplitude for the corresponding spin-flip tunneling process which results from a spin--orbit interaction with spin--orbit field $\mathbf{B}_\mathrm{SO}$ along the $y$-axis, perpendicular to the external Zeeman field $\mathbf{B}$, cf. Ref.~\cite{Stepanenko2012}. 

To describe experiments in which the PMM system is coupled to external normal metallic source and drain leads that allow for (local and nonlocal) conductance measurements, we add the following terms to our Hamiltonian
\begin{equation}
H_{rj}^{\rm res} = \sum_{k \sigma} \varepsilon_{r k \sigma } c^\dagger_{r k \sigma } c_{r k \sigma } + \left[ \sum_{k\sigma} t_{r j k \sigma}d^{\dagger}_{j\sigma}c_{r k \sigma } + \hc \right],
\label{eq:reservoirs}
\end{equation}
that describes the electronic states in lead $r$ and their coupling to the level on QD $j$.
Here, $\varepsilon_{r k \sigma }$ is the energy of an electron in level $k$ with spin $\sigma$ in lead $r$ (relative to the Fermi level) and $t_{r j k \sigma}$ parametrizes the coupling strength of that level to the localized state on the QD the lead is coupled to (for simplicity, we only include spin-conserving tunneling).
For example, for the case of a transport setup where QD $L$ is coupled to a source and QD $R$ to a drain contact, we would add $H_{SL}^{\rm res} + H_{DR}^{\rm res}$ to the Hamiltonian.
For our numerical simulations of transport experiments, we always focus on the regime where the tunnel coupling to the leads is the smallest energy scale and solve for the current with a rate-equation approach using the QMEQ package~\cite{Kirsanskas_CPC2017}.

\subsection{Spinless model}\label{sec:spinlessmodel}

When $|\varepsilon_{j\downarrow}|, |t_j|, |t_j^{{\rm SO}}|, |E_{{\rm{Z}}}^\mathrm{ABS}| \ll |\Delta|, |E_{{\rm{Z}}L,R}|$, one can neglect occupation of the excited spin state and treat the couplings between QDs $L$ and $R$ via the ABS in second order perturbation theory. The model in Eqs.~(\ref{eq:QDs})--(\ref{eq:HT}) then reduces to the effectively spinless and noninteracting model of Ref.~\cite{Leijnse_PRB2012}, sketched in Fig.~\ref{Fig1}(c) and described by
\begin{align}
H^\mathrm{spinless} = \sum_{j} \omega_j n_{j} + \left[ \chi \, d^\dagger_{L} d_{R} + \Omega \, d^\dagger_{L} d^\dagger_{R} + \hc \right]. \label{eq:ham_simp}
\end{align}
The QD orbital energy $\omega_j \approx \varepsilon_{j \downarrow}$ but is renormalized by the couplings to the ABS, while $\chi$ and $\Omega$ are the amplitudes for ECT and CAR between QDs $L$ and $R$, resulting from second order perturbation theory in $t_j$ and $t_j^{\rm{SO}}$. 

We will sometimes use the spinless model in Eq.~(\ref{eq:ham_simp}) to gain intuitive understanding or analytical results. However, it might not be experimentally possible to reach a regime where mapping onto the spinless model is appropriate, and it might not even be desirable because $|t_j|, |t_j^{\rm{SO}}| \ll |\Delta|$ implies a small gap to excited states. Importantly, it was shown in Ref.~\cite{Tsintzis2022} that one can reach a regime with high-quality PMMs also in the regime where such a mapping is not appropriate. 

In the spinless model in Eq.~(\ref{eq:ham_simp}), there is a sweet spot with a degeneracy between the lowest-energy states with even and odd fermion number parity, which is associated with having one perfect PMM localized on each QD, denoted by $\gamma_L = d_{L}^\dagger + d_{L}$ and $\tilde \gamma_R = i(d_{R}^\dagger - d_{R})$ in Fig.~\ref{Fig1}(c). This sweet spot occurs when the spin-polarized QD orbitals are at zero energy, $\omega_L = \omega_R = 0$ (i.e., they are aligned with the chemical potential of the superconductor coupling them) and when the amplitudes for CAR and ECT are equal, $\Omega = \chi$.
Ref.~\cite{Leijnse_PRB2012} proposed tuning the amplitudes for CAR and ECT through the angle between the non-collinearly polarized QD spins. In the present model, both these processes are possible because of the spin--orbit coupling, and control of the relative amplitudes can be achieved by controlling the energy of the ABS coupling the QDs because of an interference effect described in Ref.~\cite{PhysRevLett.129.267701} and studied experimentally in Ref.~\cite{Bordin2022}, see also Ref.~\cite{Tsintzis2022}.

Away from the sweet spot, the splitting between the lowest even and odd states can become finite and the corresponding fermionic mode cannot always be split up into two non-overlapping PMMs.
The effect of tuning away from the sweet spot can be understood in a more quantitative way by performing a Bogoliubov transformation to write Eq.~(\ref{eq:ham_simp}) as
\begin{align}
\begin{split}
    H^\mathrm{spinless} {} & {} = \frac{1}{2}\Big[ \sqrt{\omega_-^2 + 4 \chi^2} - \sqrt{\omega_+^2 + 4 \Omega^2}
      \Big] f^\dagger_- f_- \\
    {} & {} + \frac{1}{2}\Big[ \sqrt{\omega_-^2 + 4 \chi^2} + \sqrt{\omega_+^2 + 4 \Omega^2}
      \Big] f^\dagger_+ f_+,
\end{split}
    \label{eq:hamff}
\end{align}
up to a constant term, where $\omega_\pm = \omega_L \pm \omega_R$ and $f^\dagger_\pm$ creates a fermion in the excited (ground) state.
In terms of the Majorana operators we defined before, we can write
\begin{align}
\begin{split}
    f_- {} & {} = \tfrac{1}{2} \sin(\varphi_+) \gamma_L + \tfrac{i}{2} \sin(\varphi_-) \tilde \gamma_L \\
    {} & {} - \tfrac{1}{2} \cos(\varphi_+) \gamma_R - \tfrac{i}{2} \cos(\varphi_-)\tilde \gamma_R,
\end{split}
\label{eq:fminus}
\end{align}
where $\varphi_\pm = \tfrac{1}{2}[ \arctan(2\chi/\omega_-) \pm \arctan(2\Omega/\omega_+) ]$ and we introduced the two additional Majorana operators $\gamma_R = d_{R}^\dagger +   d_{R}$ and $\tilde \gamma_L = i(d_{L}^\dagger - d_{L})$.
We note that the two arctan functions are defined such that they yield an angle in the range $(0,\pi)$.

We see that when $\omega_L = \omega_R=0$ we have $\varphi_+ = \pi/2$ and $\varphi_- = 0$ and thus $f_- = \frac{1}{2}\gamma_L - \frac{i}{2}\tilde\gamma_R$, i.e., the lowest fermionic mode separates into two well-localized PMMs on the two QDs.
However, this does not imply that the lowest odd and even states (corresponding to having the mode $f_-$ occupied or unoccupied) must be degenerate, see Eq.~(\ref{eq:hamff}). 

\subsection{Low-energy model}\label{sec:lowE}

We also introduce an even simpler low-energy model, only including a single fermionic mode with energy $\xi$:
\begin{align}\label{eq:Hbar}
H^\mathrm{lowE} = \frac{i}{2} \xi \gamma \tilde{\gamma},
\end{align}

We choose the operators such that for unit MP, $\gamma$ has no weight on QD $R$, while $\tilde{\gamma}$ has no weight on QD $L$. In general, we let $\zeta \leq 1$ denote the relative amplitude of $\tilde{\gamma}$ on QD $L$, normalized by the amplitude of $\gamma$ on QD $L$. In the same way, we let $\tilde{\zeta} \leq 1$ denote the amplitude of $\gamma$ on QD $R$, normalized by the amplitude of $\tilde{\gamma}$ on QD $R$. Then $\zeta = 1$ ($\tilde{\zeta} = 1$) corresponds to a normal fermion on QD $L$ ($R$), while $\zeta = 0$ ($\tilde{\zeta} = 0$) corresponds to a single PMM on QD $L$ ($R$). 

The model in Eq.~(\ref{eq:Hbar}) can be seen as a low-energy approximation to both the spinful model (Sect.~\ref{sec:modelspin}) and the spinless model (Sect.~\ref{sec:spinlessmodel}). In the latter case, we define $\gamma = f_-^\dagger + f_- = \sin(\varphi_+) \gamma_L - \cos(\varphi_+)\gamma_R$ and $\tilde{\gamma} = i(f_-^\dagger - f_-) = -\cos(\varphi_-) \tilde{\gamma}_R + \sin(\varphi_-) \tilde{\gamma}_L$, and $\xi$ is the energy of the fermion annihilated by $f_-$, see Eq.~(\ref{eq:hamff}). This allows to write explicitly $\zeta = \mathrm{sin}(\varphi_-) / \mathrm{sin}(\varphi_+)$ and $\tilde{\zeta} = \mathrm{cos}(\varphi_+) / \mathrm{cos}(\varphi_-)$. 

We will use generalizations of Eq.~(\ref{eq:Hbar}) to two and three PMM systems in Sects.~\ref{sec:2DQD} and~\ref{sec:braiding} to gain intuitive understanding and compare with the physics of true topological MBSs . 

\section{Measures for Majorana quality}\label{sec:DQDMP}

When there is no clear separation between energy scales, the model in Eqs.~(\ref{eq:QDs})--({\ref{eq:HT}) does not reduce to the spinless model (\ref{eq:ham_simp}). In this case, both spin states on the QDs must be accounted for, the QD charging energy becomes important, and we cannot project out the occupation of the ABS.
Nonetheless, it was shown in Ref.~\cite{Tsintzis2022} that for a wide range of parameters, by varying $\varepsilon_{j\sigma}$ and $\varepsilon^\mathrm{ABS}_{\sigma}$ one finds two sweet spots associated with localized PMMs.
However, unlike in the spinless model in Eq.~(\ref{eq:ham_simp}), these PMMs are never perfect, even at the sweet spot where the degeneracy is (almost completely) robust to detuning of individual QD levels. In Ref.~\cite{Tsintzis2022}, the PMM quality (or closeness to perfect MBSs) was quantified through the MP defined by
\begin{eqnarray}
M_j &=& \frac{ \sum_\sigma \left( w_\sigma^2 - z_\sigma^2 \right)}{\sum_\sigma \left( w_\sigma^2 + z_\sigma^2 \right)}, \label{eq:mp}\\
w_\sigma &=& \langle o | (d_{j\sigma} + d_{j\sigma}^\dagger) |e\rangle, \\
z_\sigma &=& \langle o | (d_{j\sigma} - d_{j\sigma}^\dagger) |e\rangle,
\end{eqnarray}
where $|e\rangle$ ($|o\rangle$) is the lowest energy state with total even (odd) fermion number parity. It always holds that $-1\leq M_j  \leq 1$. It should be noted that this definition of the MP assumes real wavefunctions, which we can always choose because the Hamiltonians in Eqs.~({\ref{eq:QDs})--(\ref{eq:HT}}) are real; for complex wavefunctions there is a phase degree of freedom in the definition of $w_\sigma$ and $z_\sigma$ which requires some extra care.

The MP (\ref{eq:mp}) can be calculated anywhere in parameter space, but we would only really call the states PMMs if we have a combination of high (close-to-unity) MP and a \mbox{(quasi-)}degeneracy between the even and odd  parity ground states.
$|M_j|=1$ means that only a single Majorana operator that switches between these even and odd parity ground states has any weight on QD $j$. In the different experiments proposed and investigated below, we will study effects of having a low MP even at the best possible sweet spot (resulting from a too low $E_{{\rm{Z}}j}$ and $U_j$~\cite{Tsintzis2022}). We will also consider (intentional and unintentional) deviations from the sweet spot, which can break the even--odd degeneracy and/or reduce the MP.

In the spinless model described in Sect.~\ref{sec:spinlessmodel}, we can use Eq.~(\ref{eq:fminus}) to find explicit expressions for the MP
\begin{align}
    M_L = {} & {} \frac{\sin(\varphi_+)^2 - \sin(\varphi_-)^2}{\sin(\varphi_+)^2 + \sin(\varphi_-)^2} \label{eq:mp1_L}\\
    = {} & {}  \frac{-4\chi\Omega}{\omega_+ \omega_- - \sqrt{(\omega_-^2+4\chi^2)(\omega_+^2 + 4\Omega^2)}},\label{eq:mp_simp_L}
\end{align}
and
\begin{align}
     M_R = {} & {} \frac{\cos(\varphi_+)^2 - \cos(\varphi_-)^2}{\cos(\varphi_+)^2 + \cos(\varphi_-)^2} \label{eq:mp1}\\
    = {} & {}  \frac{-4 \chi \Omega}{\omega_+ \omega_- + \sqrt{(\omega_-^2+4\chi^2)(\omega_+^2 + 4\Omega^2)}}.\label{eq:mp_simp}
\end{align}
We note that these expressions simply quantify how purely real ($\gamma_j$) or imaginary ($\tilde \gamma_j$) the component of the ground-state wavefunction on QD $j$ is, perfect MP corresponding to $M_L = \pm 1, M_R = \mp 1$.

Within the low-energy model introduced in Sect.~\ref{sec:lowE}, it seems reasonable to define the MP as 
\begin{equation}\label{eq:lowEMP}
    M_L \approx \frac{1 - \zeta^2}{1 + \zeta^2}, \; \; M_R \approx -\frac{1 - \tilde{\zeta}^2}{1 + \tilde{\zeta}^2}.
\end{equation}
This clearly agrees with Eqs.~(\ref{eq:mp1_L}) and~(\ref{eq:mp1}) if we derive the low-energy model from the spinless model, but we emphasize that for relatively high MP (small $\zeta$) Eqs.~(\ref{eq:Hbar}) and~(\ref{eq:lowEMP}) hold more generally, i.e., also when the PMMs have some weight on the ABS and on the excited spin state. When investigating coupled PMM systems and braiding in Sects.~\ref{sec:2DQD} and~\ref{sec:braiding}, we will see that, assuming that we have tuned the system to even--odd degeneracy, $\zeta$ and $\tilde{\zeta}$ indeed quantify how the results deviate from those expected for topological MBSs. 

We emphasize that $M_j$ is normalized to the total weight of the wavefunction on the QD under consideration.
This suggests that one can have $|M_j| = 1$ even with only negligible average occupation of QD $j$.
This can be clearly seen in the spinless model, where, for example, in the limit $\omega_R \to \infty$ and $\omega_L=0$ one finds for $\chi = \Omega$ that the even and odd ground states are the trivial states $\ket{0}$ and $d^\dagger_{L}\ket{0}$;
however, Eq.~(\ref{eq:mp_simp}) yields in this case $M_L = 0$ for QD $L$, but $M_R = -1$ for the empty QD $R$.
The quantities $M_{L,R}$ are thus not very good measures for the ``Majorananess'' of the ground state wavefunction as a whole.
They are, however, useful to consider when one is interested in coupling multiple PMM systems, to form qubits and perform fusion or braiding experiments:
As discussed in Sect.~\ref{sec:coupledDQDs}, when two PMMs are connected via two QDs $AR,BL$ that have $|M_{AR}| = |M_{BL}| = 1$, the resulting coupling in the low-energy sector can be written in the form $H^\mathrm{lowE}_{AB} = \frac{i}{2}\lambda_{A B} \gamma_A \gamma_B$; reduced weights on the coupled QDs reduce the effective coupling strength $\lambda_{A B}$ but will not introduce any undesired finite coupling matrix elements between other Majorana components of the low-energy modes.

We note that the product of the two polarizations $-M_L M_R$ or their weighted sum $\frac{1}{2}(M_L-M_R)$ could be used as a measure for the Majorana quality of the whole PMM system, both measures reaching 1 in the ideal case.

In case one is interested in a measure that takes into account the relative weight of the mode on the outer QDs, one could also consider the quantity
\begin{align}
\begin{split}
    M_\mathrm{G} = {} & {} \tfrac{1}{2}\big[ \cos(\varphi_+)^2 - \cos(\varphi_-)^2 \big]^2\\
     + {} & {} \tfrac{1}{2}\big[ \sin(\varphi_+)^2 - \sin(\varphi_-)^2 \big]^2
\end{split}\\
= {} & {} \frac{16 \chi \Omega}{(\omega_-^2+4\chi^2)(\omega_+^2 + 4\Omega^2)},\label{eq:mbar}
\end{align}
written here in terms of the spinless model (one could straightforwardly construct an equivalent expression for the spinful case).
This quantity satisfies $0 \leq M_\mathrm{G} \leq 1$ and only reaches its maximum when $f_- = \pm \frac{1}{2}\gamma_L \mp \frac{i}{2}\tilde \gamma_R$ or $f_- = \mp \frac{1}{2} \gamma_R \pm \frac{i}{2} \tilde \gamma_L$, i.e., when the lowest-energy mode is a perfect MBS with all its weight on the two outer QDs.
This $M_\mathrm{G}$ is thus more closely related to a global Majorana quality of the full wavefunction corresponding to the lowest mode.
However, in the spinful model the wavefunction will also have some weight on the ABS and a generalization of $M_\mathrm{G}$ will thus rarely approach 1. The same will hold for any more complex model with additional degrees of freedom. A low value of $M_G$ could indicate a strong mixing of the Majorana components of the wavefunction, but could also be merely related to a small total weight of the state on the outer QDs.

For the experiments considered in Sects.~\ref{sec:2DQD} and~\ref{sec:braiding}, only the Majorana quality in the parts of the PMM system that are coupled to other components matters. This is quantified by the normalized local $M_j$, and we therefore focus on this quantity in the following.

\section{Detecting Majorana states}\label{sec:detection}

\subsection{Tunneling spectroscopy}\label{sec:conductance}

The most straightforward way to probe the low-energy physics in a PMM system is to connect it via tunneling barriers to metallic reservoirs, as sketched in Fig.~\ref{FigPC}(a), and perform tunneling spectroscopy.
In such a setup the full differential conductance matrix $G_{jk} = dI_j / dV_k$, with $j,k = L,R$, can be accessed by varying the voltages $V_k$ applied to the reservoirs and monitoring the currents $I_j$ flowing into the two sides of the PMM system.
A peak in the local differential conductances $G_{jj}$ at small bias voltage signals a degeneracy of the lowest even and odd states.
Mapping out the structure of this degeneracy as a function of the QD orbital energies, which can be controlled via electrostatic gates, can provide information about the ratio of CAR and ECT and could thus serve as a guidance for tuning towards points in parameter space with large MP~\cite{Tsintzis2022,Dvir2023}.
A measurement of the nonlocal differential conductances $G_{LR}$ and $G_{RL}$ provides additional information about the detailed structure of the bound state involved in the transport, including its ``BCS charge'' distribution over the two QDs~\cite{Danon_non-local_2020}, such that a sign change of $G_{LR}$ and $G_{RL}$ coinciding with the degeneracy can provide extra evidence for a bound state with high MP~\cite{Dvir2023}.

\subsection{Coupling to an extra quantum dot}\label{sec:pradaclarke}

Another method to assess the quality of the MBSs and indirectly probe their MP is to add one extra QD to the setup and measure the conductance of the combined QD--PMM setup, see Fig.~\ref{FigPC}(b).
While direct local tunneling spectroscopy on the bare PMM system can only reveal the presence of a low-energy state that has significant weight on one end of the system, the addition of the extra QD makes the level structure of the combined setup sensitive to the exact distribution of the state over the whole PMM part of the system, thus providing insight into the Majorana quality of the state through straightforward local conductance measurements only.
In the search for MBSs in proximitized nanowires, the addition of such an extra QD indeed yielded features in the low-energy part of the spectrum that have been interpreted as signatures of the absence or presence of low-energy modes corresponding to localized MBSs~\cite{deng2016majorana, Prada_PRB2017, Clarke_PRB2017, deng2018nonlocality}.

To investigate this setup in more detail for the PMM system, we thus add an extra QD ``$D$'' to our model, which we account for by adding to the Hamiltonian
\begin{align}
\begin{split}
H_{\rm D} {} & {} = \sum_\sigma \varepsilon_{D\sigma} n_{D\sigma}
+ U_D n_{D\uparrow} n_{D\downarrow} 
\\
{} & {} + \sum_\sigma \Big[ t_{D} d_{D \sigma}^\dagger d_{L\sigma}  + t^\mathrm{SO}_{D} s_\sigma d_{D \sigma}^\dagger d_{L\bar{\sigma}} + \hc \Big],
\end{split}
\end{align}
where $d^\dagger_{D\sigma}$ creates an electron with spin $\sigma$ on the extra QD $D$.
QD $D$ is only tunnel coupled to QD $L$ of the PMM system [see Fig.~\ref{FigPC}(b)] and the orbital on the QD $D$ has the (spin-dependent) single-particle energy $\varepsilon_{D\sigma} = \varepsilon_D \pm E_{{\rm Z}D}/2$ and an on-site Coulomb charging energy of $U_D$.
For simplicity, we did not include inter-QD electrostatic interaction between QDs $D$ and $L$, which could in principle be significant since they are not separated by a superconducting element that screens the interaction.
We will assume here that such interactions are being screened by additional metallic components or actively compensated for in the gating of the device.

This combined QD--PMM system is then connected to a source and a drain lead, as indicated in Fig.~\ref{FigPC}(b).
We include this into our model by adding the terms $H^{\rm res}_{LD} + H^{\rm res}_{RR}$ to the Hamiltonian [see Eq.~(\ref{eq:reservoirs})] and numerically calculate the current through the system, see Sec.~\ref{sec:modelspin}.

\begin{figure}[t!] \centering
\includegraphics[width=1\linewidth]{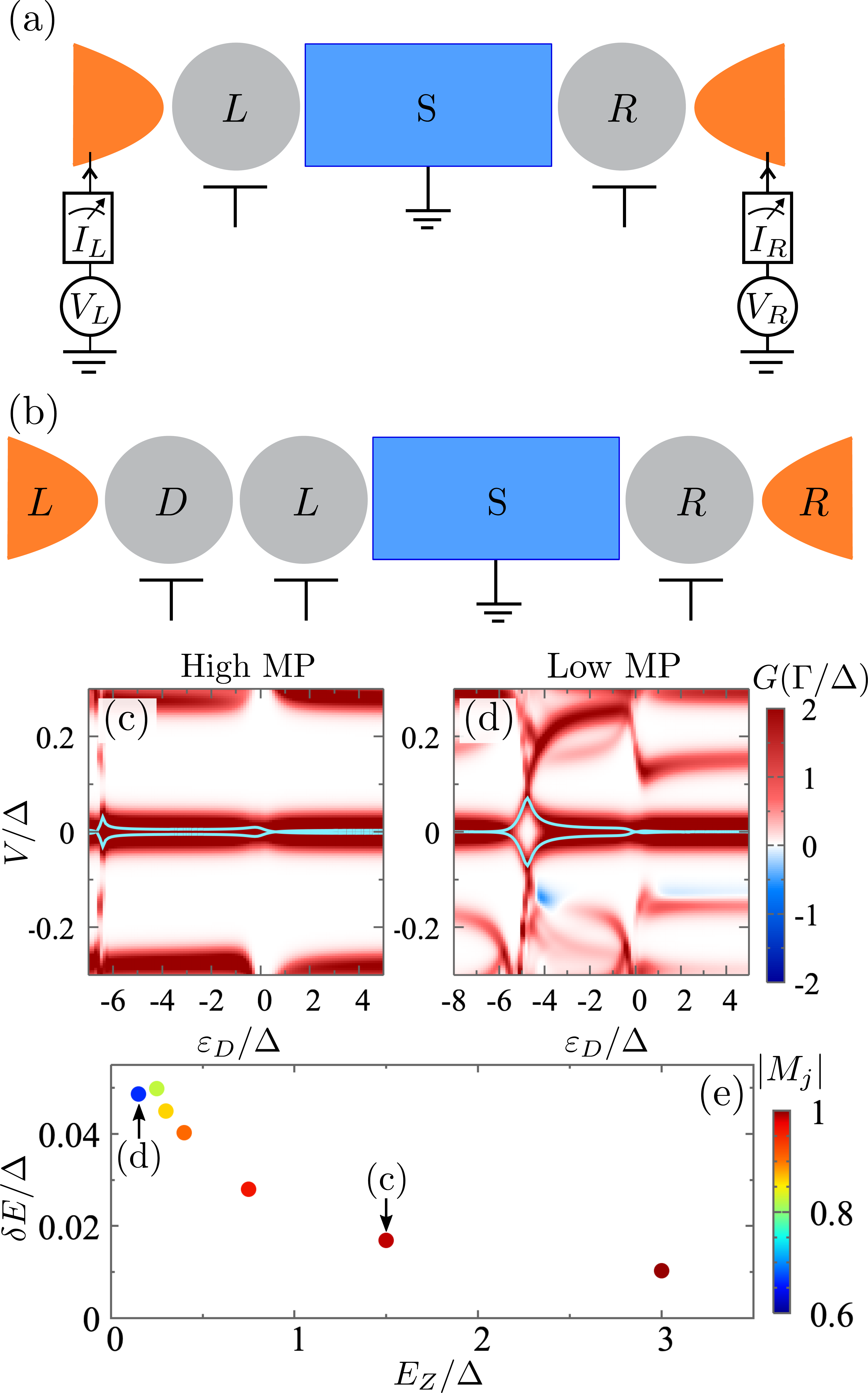}
\caption{
(a) Transport setup that allows for measurement of the full differential conductance matrix of a PMM system, such as performed in Ref.~\cite{Dvir2023}.
(b) QD--PMM system embedded in a similar transport setup as in (a).
(c) Local differential conductance ($G=G_{RR}$) of the QD--PMM system for a high-MP sweet spot as a function of bias voltage $V/2 = V_L = -V_R$ and QD orbital energy $\varepsilon_D$. The solid light blue lines show the difference between the lowest even and odd states obtained from numerical diagonalization of the spinful Hamiltonian. 
(d) Same as (c) but for a low-MP sweet spot.
(e) Maximum splitting between the even and odd ground states when sweeping $\varepsilon_D$ as a function of the Zeeman splitting on all the QDs, for parameters corresponding to different MP values (indicated by the dot color).
In all calculations we used (all in units of $\Delta$) $U_L=U_R=U_D=5$ and $E_{\rm Z}^{\rm ABS}=0$.
All the tunnel couplings between the QDs are chosen equal to $t_j=0.5$ and the spin--orbit tunnel couplings are set to $t_j^\mathrm{SO} = 0.1$.
To change the MP we varied $E_{{\rm Z}L,R}$ (with $E_{{\rm Z}D}=E_{{\rm Z}L,R}$) and adjusted $\varepsilon_{L,R}$ to find a sweet spot.
In (c) (high MP, $|M_{L,R}| \approx 0.986$) we used $E_{{\rm Z}L}=E_{{\rm Z}R}=E_{{\rm Z}D}=1.5$, $\varepsilon_L = \varepsilon_R=-0.154$, $\varepsilon^\mathrm{ABS}=-0.329$.
In (d) (low MP, $|M_{L,R}| \approx 0.661$) we used $E_{{\rm Z}L}=E_{{\rm Z}R}=E_{{\rm Z}D}=0.15$, $\varepsilon_L = \varepsilon_R=-0.306$, and $\varepsilon^\mathrm{ABS}=-0.564$.
}\label{FigPC}
\end{figure}

Figures~\ref{FigPC}(c,d) show the calculated local differential conductance $G_{RR}$ as a function of the symmetrically applied bias voltage $V/2=V_L = -V_R$ and QD detuning $\varepsilon_D$.
We plot the conductance in units of $\Gamma/\Delta$, where $\Gamma = 2\pi t_{rjk\sigma}^2 \nu_{\rm res}$ is the tunnel rate to the reservoirs, with $\nu_{\rm res}$ the density of states of the reservoirs, which, together with the tunnel coupling coefficients $t_{rjk\sigma}$, is assumed to be energy- and spin-independent and is set the same for the two sides of the system.  Within the rate equation, $\Gamma$ is just a prefactor and the conductance plotted in units of $\Gamma/\Delta$ is independent of $\Gamma$. However, the rate equation is only valid under the assumption $\Gamma \ll T$, where we take the thermal energy to be $T= \Delta/200$.  All other parameter values are specified in the caption of Fig.~\ref{FigPC}. 

In all the panels, the PMM system is tuned to a sweet spot, i.e., a point in the parameter space $(\varepsilon_L,\varepsilon_R, \varepsilon^\mathrm{ABS})$ where the lowest even and odd state are degenerate while the MP is maximized, cf.~Ref.~\cite{Tsintzis2022}.
Figure~\ref{FigPC}(c) shows the case where the sweet spot has relatively high MPs, $|M_R| = |M_L| = 0.986$, whereas in Fig.~\ref{FigPC}(d) we have lower $|M_R| = |M_L| = 0.661$, due to a ten times lower Zeeman splitting.
The solid light blue lines indicate the energy difference between the lowest even and odd states, found from numerically diagonalizing the spinful Hamiltonian.

Similar to the case with MBSs in nanowires~\cite{deng2016majorana, Prada_PRB2017, Clarke_PRB2017, deng2018nonlocality}, we see that the case with high MP distinguishes itself by showing much smaller splitting of the zero-bias peak as a function of the QD level $\varepsilon_D$, a feature that is indeed clearly visible in the local conductance data.
The intuitive picture is that if QD $D$ only couples to a single PMM, it cannot break the ground state degeneracy and level crossings will thus not produce significant splitting.
In general, we note that the line shapes we observe close to the level crossings strongly resemble those observed in experiments~\cite{deng2016majorana,deng2018nonlocality}, which were called ``bowtie'' and ``diamond'' patterns~\cite{Prada_PRB2017,Clarke_PRB2017}.

To corroborate the connection between the observed level splitting at the crossing and the MP of the PMM system, we show in Fig.~\ref{FigPC}(e) the maximal even--odd ground-state splitting $\delta E$ around the level crossings as a function of the Zeeman splitting on the three QDs.
For each point, we tuned the PMM system to a sweet spot, where the even and odd ground states are degenerate and the MP is maximal.
The colors of the dots indicate the MP, which ranges from 0.661 for a Zeeman splitting of $E_{\rm Z}= 0.15\,\Delta$ to 0.995 for $E_{\rm Z} = 3\,\Delta$.
The dependence of the MP on the Zeeman splitting is well know~\cite{Tsintzis2022}; additionally we see a clear correlation between the MP and the maximal $\delta E$ when sweeping $\varepsilon_D$, with $\delta E$ converging to 0 for $|M_L|=1$.

These results clearly indicate that the addition of an extra QD can facilitate probing the Majorana quality of the low-energy states in a PMM system, using local tunneling spectroscopy only.

\subsection{Quantum capacitance and Majorana polarization}\label{sec:capacitance}

Another method to access the Majorana quality of a PMM system could be to monitor the quantum capacitance of the lowest even and odd state with respect to one of the gate voltages that control the QD levels.
In the spinless model, the even (odd) charge sector is spanned by the two basis states $\ket{00} = \ket{0}$ and $\ket{11} = d^\dagger_{L}d^\dagger_{R}\ket{0}$ ($\ket{10} =  d^\dagger_{L}\ket{0}$ and $\ket{01} = d^\dagger_{R}\ket{0}$).
At the sweet spot, where $\omega_L=\omega_R=0$, the lowest even and odd eigenstates become $\frac{1}{\sqrt 2}(\ket{00} - \ket{11})$ and $\frac{1}{\sqrt 2}(\ket{10}-\ket{01})$, respectively, together defining a fermionic mode that is comprised of two perfect MBSs.
Both these states are equal superpositions of different charge states, which is indeed a crucial ingredient to make them indistinguishable by local charge measurements, a characteristic property related to the nonlocal nature of MBSs.
However, this also implies that at the sweet spot the charge distribution of both states is most sensitive to small changes in the on-site potentials, suggesting a local maximum of the quantum capacitances $ C_{pj} = d^2 E_p/d \omega_j^2$, where $p = e,o$ denoting the parity of the state, and $j = L,R$.

The simplicity of the spinless model allows us to straightforwardly find analytic expressions for the capacitances,
\begin{align}
    C_{ej} = {} & {}  -\frac{2 \Omega^2}{\big( \omega_+^2 + 4\Omega^2\big)^{3/2}}, \\
    C_{oj} = {} & {}  -\frac{2 \chi^2}{\big( \omega_-^2 + 4 \chi^2\big)^{3/2}},
\end{align}
which are equal for $j = L,R$ and are indeed both maximal at the point $\omega_{L,R}=0$.
In fact, we see that in the spinless model there is a direct connection with one of the MPs defined in Eq. \eqref{eq:mbar},
\begin{align}
    C_{ej} C_{oj} = \frac{M_\mathrm{G}^{3/2}}{16 \chi \Omega}.\label{eq:cc}
\end{align}
A local maximum of $C_{ej} C_{oj}$ as a function of tuning parameters can thus be interpreted as a signal of high MP.

To find the sweet spot in an experiment, one could first find the manifold in parameter space that yields a vanishing even--odd splitting, using local tunneling spectroscopy.
After that, one could monitor the capacitance of the QDs of the PMM system, via RF reflectometry (see, for example, Ref.~\cite{Petersson2010, Lambert2016, Vigneau2023}), while allowing for tunneling events that change the parity of the system, or just rely on randomly occurring quasiparticle poisoning processes to switch between the even and odd ground states.
From the telegraph signal measured one can extract a measure for both $C_{ej}$ and $C_{oj}$. The product $C_{ej} C_{oj}$ is expected to have a maximum at the Majorana sweet spot for zero energy splitting, which can be used to tune the energy of the QDs. 

\begin{figure}[t!] \centering
\includegraphics[width=1\linewidth]{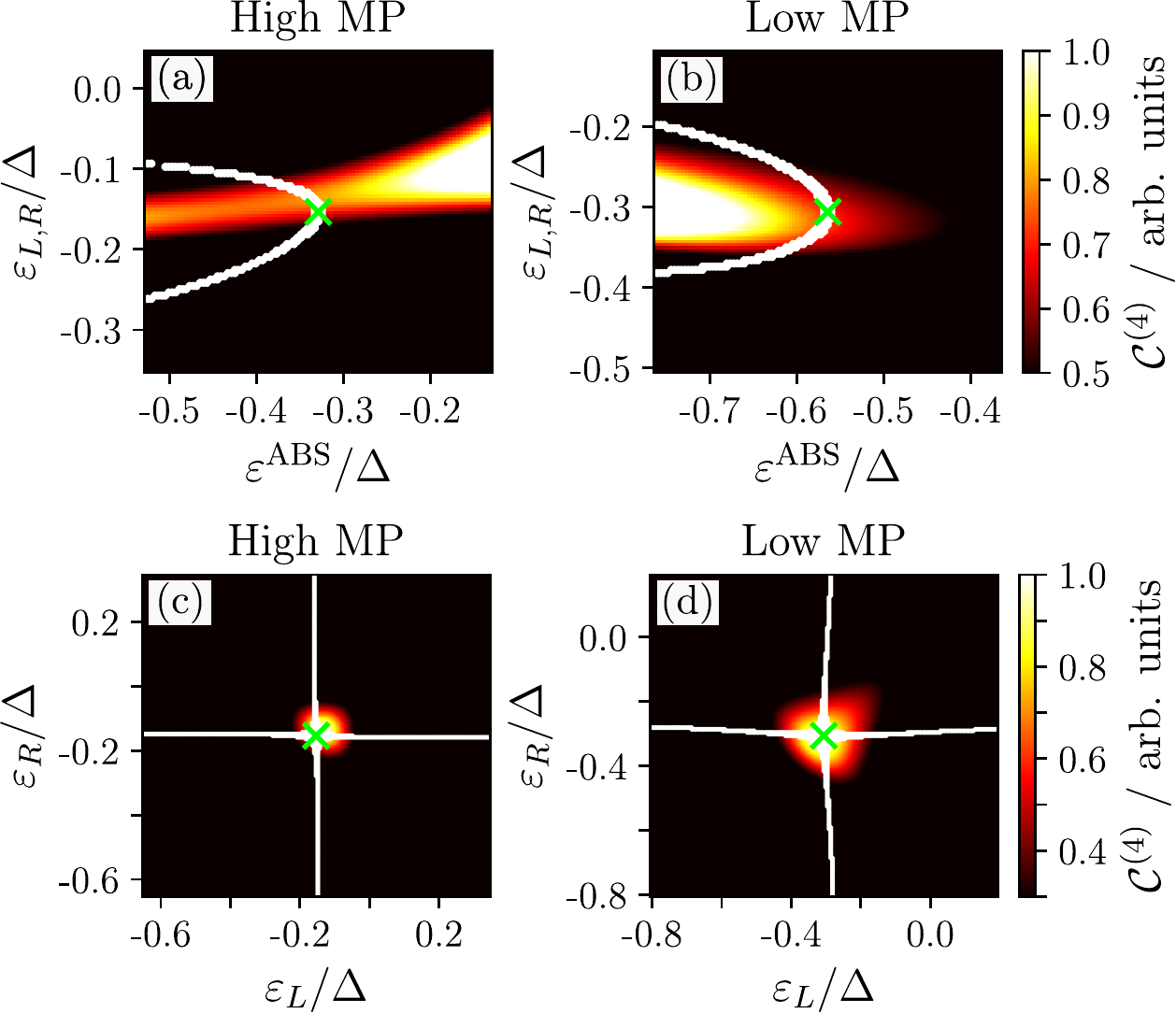}
\caption{Product of the four capacitances ${\cal C}^{(4)}$ as a function of the gate-tunable parameters (a,b) $\varepsilon_{L} = \varepsilon_R$ and $\varepsilon^\mathrm{ABS}$ or (c,d) $\varepsilon_L$ and $\varepsilon_R$.
In (a,c) we used the same parameters as in Fig.~\ref{FigPC}(c) and in (b,d) we used the same parameters as in Fig.~\ref{FigPC}(d).
The green crosses mark the locations of the local sweet spot, where the even--odd splitting vanishes and the local $M_{L,R}$ are maximized.
White lines indicate the trajectory of the ground state degeneracy in parameter space.
}\label{FigCap}
\end{figure}

We now explore to what extent this relationship holds in the spinful interacting model, with finite on-site Zeeman splittings.
The capacitances are now defined as $C_{pj} = d^2 E_p/d \varepsilon_j^2$ and are in general no longer symmetric in $j = L,R$.
In Fig.~\ref{FigCap} we thus plot the product of the four capacitances ${\cal C}^{(4)} = C_{eL} C_{eR} C_{oL} C_{oR}$ and assess the correlation between local maxima of this quantity and the location of the sweet spot.
Figures.~\ref{FigCap}(a,b) show ${\cal C}^{(4)}$ as a function of $\varepsilon_{L} = \varepsilon_R$ and $\varepsilon^\mathrm{ABS}$.
In Fig.~\ref{FigCap}(a) we used a high Zeeman splitting of $E_{{\rm Z} L,R} = 1.5\,\Delta$, resulting in a sweet spot with a relatively high MP, whereas we used $E_{{\rm Z} L,R} = 0.15\,\Delta$ in Fig.~\ref{FigCap}(b), yielding a low-MP sweet spot (see the caption of Fig.~\ref{FigPC} for all parameters used).
In both plots we indicate the location of the even--odd degeneracy with a white line and the location of the sweet spot is marked with a green cross.
In both cases, the sweet spot indeed coincides with a maximum of ${\cal C}^{(4)}$ along the degeneracy.
Figs.~\ref{FigCap}(c,d) show the same two situations, now as a function of independent $\varepsilon_L$ and $\varepsilon_R$, with fixed $\varepsilon^\mathrm{ABS} = -0.329\,\Delta$ (c) and $\varepsilon^\mathrm{ABS} = -0.564\,\Delta$ (d).
In this case the correlation between maximum ${\cal C}^{(4)}$ and MP is even more clear than in Figs.~\ref{FigCap}(a,b).

This indeed suggests that local tunneling spectroscopy combined with quantum capacitance measurements could provide enough information to identify the sweet spots in parameter space.
The method only relies on finding the maximum of ${\cal C}^{(4)}$ within the even--odd ground-state degeneracy manifold. This tuning method provides more information than transport measurements alone, which can only determine conditions for ground state degeneracies, as it is insensitive to details such as QD level arms and cross-capacitances between different gates. However, quantum capacitance measurements cannot distinguish between high and low MP sweet spots. To distinguish them, experiments probing PMM's nonabelian properties are required, discussed in Sec. \ref{sec:braiding}.

\section{Initialization and readout}\label{sec:initialization}

To go beyond transport spectroscopy, approach Majorana qubits and eventually nonabelian physics, the first step is to develop the ability to initialize and read out the state associated with a pair of PMMs. Here, the PMM's lack of topological protection turns into an advantage. The ground state is singly degenerate everywhere, except for fine-tuned situations that include the sweet spot where well-separated PMMs appear. Thus, to initialize the system we can simply shift the orbitals of both QDs away from the sweet spot (the degeneracy is not lifted for small shifts of the orbital of only a single QD, although the PMM localization is affected~\cite{Leijnse_PRB2012}). 

Figures~\ref{fig:init_readout}(a,b) show the energy differences $\delta E_n$ between the lowest excited states $n = 1, 2, 3$ and the ground state as a function of the detuning from the sweet spot (purple lines), for a sweet spot with high MP in (a) and low MP in (b) (using the same parameters as in Fig.~\ref{FigPC}). Here, we have chosen a symmetric detuning, $\delta \varepsilon_L = \delta \varepsilon_R$, where $\delta \varepsilon_j$ is the deviation of $\varepsilon_j$ from the sweet spot value. For symmetric detuning the ground state is even, while an asymmetric detuning ($\delta \varepsilon_L = - \delta \varepsilon_R$) would give an odd ground state. Initialization is most easily done by detuning to a point where the even--odd splitting is larger than the thermal energy and waiting for a time longer than the quasiparticle poisoning time. 

\begin{figure}[t] \centering
\includegraphics[width=1\linewidth,trim={0cm 0cm 0cm 0cm},clip]{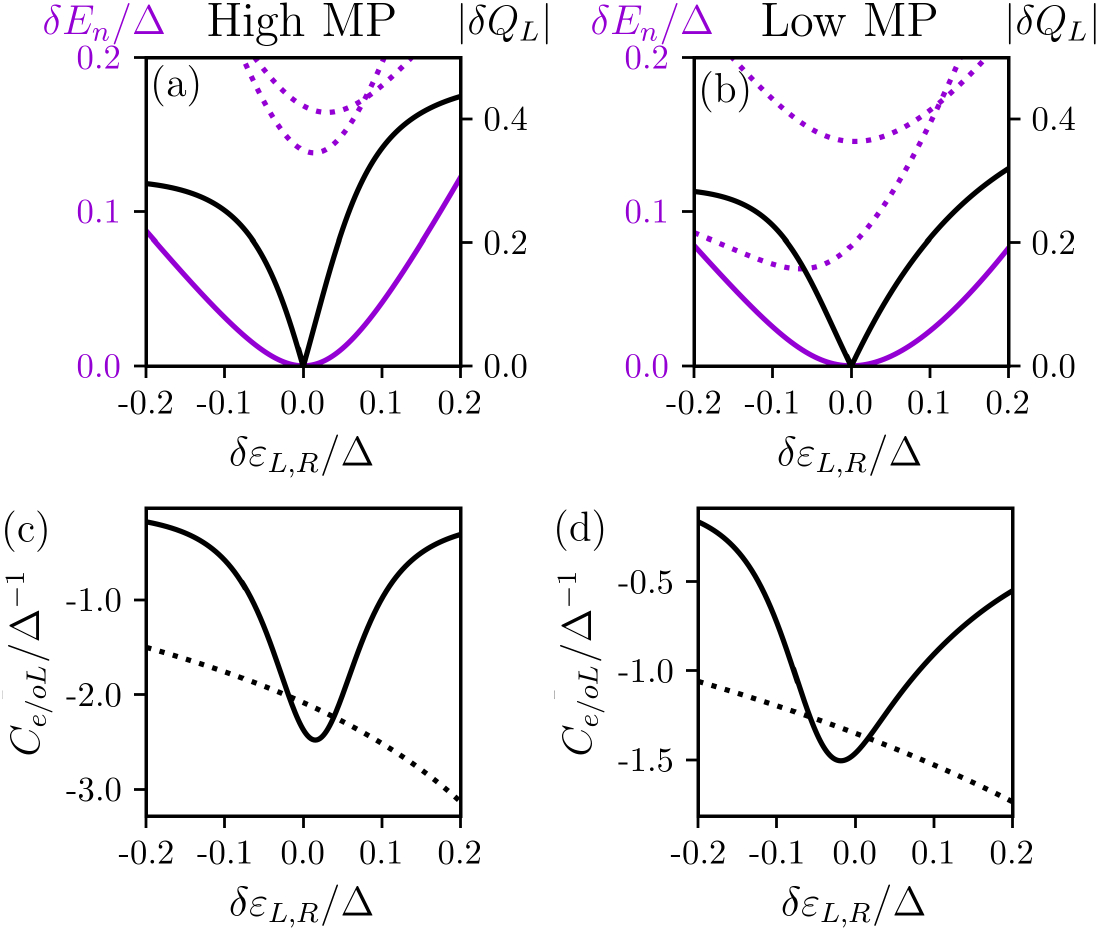}
\caption{(a) and (b) Excitation energies $\delta E_n$ (purple lines) and absolute value of the charge difference $|\delta Q_L|$ between even and odd ground states on the left QD (black lines) as a function of detuning both QD orbitals symmetrically ($\varepsilon_L = \varepsilon_R$) away from the sweet spot. (c) and (d) Similar to (a) and (b), but showing instead the quantum capacitance associated with the left QD of the even ground state ($C_{eL}$, solid) and odd ground state ($C_{oL}$, dotted). (a) and (c) correspond to parameters where the MP is high at the sweet spot, while (b) and (d) correspond to parameters where the MP is low at the sweet spot [same parameters as in Fig.~\ref{FigPC}(c,d)].}
\label{fig:init_readout}
\end{figure}

The conceptually simplest way to read out the state encoded in the two MBSs is through charge detection on one (or both) QDs, similar to what is done for spin qubits~\cite{Elzerman2004, Reilly2007, Barthel2009, Volk2019, Liu2021a}. This does not give a signal at the sweet spot, where the charge on both QDs is equal in the even and odd states. But as the QD orbitals are detuned a substantial charge difference develops, see the black lines in Figs.~\ref{fig:init_readout}(a,b) which show $|\delta Q_L| = | \langle e | Q_L | e \rangle - \langle o | Q_L | o \rangle |$. This parity-to-charge conversion is similar to spin-to-charge conversion used for single-shot readout of spin qubits in double QDs~\cite{Barthel2009}.

Another option is to read out the state based on a measurement of the quantum capacitance introduced in Sect.~\ref{sec:capacitance}, see, for example, Refs.~\cite{Petersson2010, Lambert2016, Vigneau2023}. For readout, we want to measure the quantum capacitance of an individual QD which is plotted in Figs.~\ref{fig:init_readout}(c) and (d) for a high- and low-MP sweet spot, respectively. At the sweet spot, $C_{ej} \approx C_{oj}$, the difference approaches zero as MP approaches unity, and is exactly zero at the sweet spot in the spinless model in Eq.~(\ref{eq:ham_simp}). But away from the sweet spot, the quantum capacitances differ substantially between the lowest even and odd states. Just as for charge detection, we can therefore use a quantum capacitance measurement for readout if we first detune the system away from the sweet spot.

We note that it is also possible to construct a setup that allows reading out the state while keeping the system at the sweet spot~\cite{Leijnse_PRB2012, Liu2022_PMMfusion}. However, adding detectors capable of readout at the sweet spot not only complicates the device and measurements (requiring either measurements sensitive to charge fluctuations or capacitive coupling to both QDs) but also necessarily introduces a decoherence mechanism at the sweet spot (which will be relevant when we consider coupled PMM systems below). We, therefore, believe that it is better to use simpler readout schemes, such as discussed above, which are only sensitive to the state encoded in the PMMs away from the sweet spot.

Using the methods for initialization and readout discussed above, we propose that an important first experiment is to measure the lifetime of the parity of a single PMM system. This will limit the lifetime of a PMM qubit and be a limiting factor for any experiment probing the nonabelian nature of the low-energy states. Away from the sweet spot, continuous charge detection or quantum capacitance measurements should exhibit jumps in the readout signal when quasiparticle poisoning switches the system between even and odd parity, thus revealing the rate for such processes. This requires staying close enough to the sweet spot that the even and odd ground states are not separated by much more than the thermal energy. The magnitude of the jumps in the signal should be reduced when approaching the sweet spot. To measure the parity lifetime at the sweet spot, one could instead initialize the system away from the sweet spot, tune back to the sweet spot for some waiting time, followed by readout away from the sweet spot. 


Finally, we note that the physics discussed in this section is to a large extent independent of the MP at the sweet spot. Both readout methods discussed above (charge detection and quantum capacitance measurements) work almost equally well for a high-MP sweet spot [Figs.~\ref{fig:init_readout}(a) and (c)] as for a low-MP sweet spot [Figs.~\ref{fig:init_readout}(b) and (d)]. There is no qualitative difference in the spectra between high- and and low-MP sweet spots [compare Figs.~\ref{fig:init_readout}(a) and (b)], although the gap to excited states tends to be lower for lower MP. There is also no reason to expect drastic differences in the quasiparticle poisoning times between PMM systems with high- and low-MP sweet spots. The difference between high and low MP will, however, be clear when coupling two PMM systems, as will be discussed in Sect.~\ref{sec:2DQD}, and will be of crucial importance for the braiding protocols investigated in Sect.~\ref{sec:braiding}. 

\section{Coherent Majorana operations}\label{sec:2DQD}
In the previous sections, we have described ways to characterize a PMM system, where information is encoded in the parity degree of freedom. However, it is usually not possible to create superpositions of the even and odd parity states encoded in a single pair of PMMs (or topological MBSs). Therefore, a Majorana qubit should be encoded in four PMMs, which can be hosted in two PMM systems with a fixed total parity~\cite{Bravyi2006, Leijnse_Review2012}. 

\subsection{Coupled double quantum dots}\label{sec:coupledDQDs}
We consider two different realizations of the coupling between two PMM systems, as shown in Figs.~\ref{fig:twoDQDs}(a) and (d). We denote the two PMM systems with $A$ and $B$, each of which is described by a Hamiltonian as in Eqs.~(\ref{eq:QDs})--(\ref{eq:HT}).

The first way to couple the two PMM systems is by direct coupling between the two closest QDs, described by
\begin{eqnarray}
H_{AB} &=& \sum_{\sigma} \Big[ t_{A B} d_{AR \sigma}^\dagger d_{A R\sigma}  + t_{A B}^\mathrm{SO} s_\sigma d_{AR \sigma}^\dagger d_{BL \bar{\sigma}}+ \hc \Big] \label{eq:HTAB} \nonumber \\
&+& U_{AB} n_{AR} n_{BL},
\end{eqnarray}
where $t_{A B}$ ($t_{A B}^\mathrm{SO}$) is the amplitude for spin-conserving (spin--orbit-induced spin-flip) tunneling between the right QD in system $A$ and the left QD in system $B$ and $U_{AB}$ is the inter-QD Coulomb charging energy between those QDs with $n_{AR} = \sum_\sigma n_{A R\sigma}$, $n_{BL} = \sum_\sigma n_{BL\sigma}$. A finite $U_{AB}$ will induce correlations between the charge on the innermost QDs and prevent reaching a sweet spot in any of the individual PMM systems even with the tunnel coupling switched off, and we will in the following assume $U_{AB} = 0$. Because, in contrast to the individual PMM systems, there is no natural screening by a superconducting coupler, achieving $U_{AB} = 0$ will require some design effort, just as for the extra QD in Sect.~\ref{sec:pradaclarke}.

The complication with an inter-QD charging energy can be circumvented by coupling the two PMM systems via an additional superconducting segment. An additional potential advantage with this coupling is that the whole system can be turned into a four-site Kitaev chain. We describe this superconductor with a Hamiltonian analogous to Eq.~(\ref{eq:ABS}), while the couplings are described by a Hamiltonian analogous to Eq.~(\ref{eq:HT}}):
\begin{align}
\begin{split}H_{ACB} &= \sum_{\sigma} s_\sigma \Big[t^\mathrm{SO}_{AC} d_{AR \sigma}^\dagger c_{C \bar{\sigma}} + t^\mathrm{SO}_{BC}  c_{C \sigma}^\dagger d_{BL \bar{\sigma}} \Big]  \\
&+ \sum_\sigma \Big[ t_{AC} d_{AR \sigma}^\dagger c_{C \sigma} + t_{BC}  c_{C \sigma}^\dagger d_{BL \sigma} \Big]   + \hc,\label{eq:tildeHTAB}
\end{split}
\end{align}
where $t_{mC}$ ($t^\mathrm{SO}_{mC}$) is the amplitude for spin-conserving (spin-flip) tunneling between PMM system $m=A,B$ and the connecting superconductor ($C$) that has electron annihilation operator $c_{C\sigma}$. 

With both ways to couple the two PMM systems, the phase differences between the superconductors becomes important, which can be transformed into phases in the couplings $t_{A B}$ or $t_{AC}, t_{BC}$. In the results presented in Figure~\ref{fig:twoDQDs} we have taken all such couplings to be real, which in general will require connecting the different superconductors in a loop to allow phase control. This control is not crucial for the physics discussed here, but as we will see below it is advantageous for suppressing undesired PMM couplings for low-MP sweet spots.

We consider a setup where the couplings between PMM systems $A$ and $B$ (via either of the two mechanisms described above) can be controlled via gate voltages. If PMM systems $A$ and $B$ are both tuned to a sweet spot and the coupling between them is switched off, the ground state is fourfold degenerate. We can choose to operate the system within the subspace with total even or odd parity, spanned by $|ee\rangle = |e\rangle_A |e\rangle_B, |oo\rangle$ and $|eo\rangle, |oe\rangle$ respectively. Transitions between the total even and total odd parity subspaces can only happen via quasiparticle poisoning.

\begin{figure}[t] \centering
\includegraphics[width=1\linewidth,trim={0cm -1cm 0cm 0cm},clip]{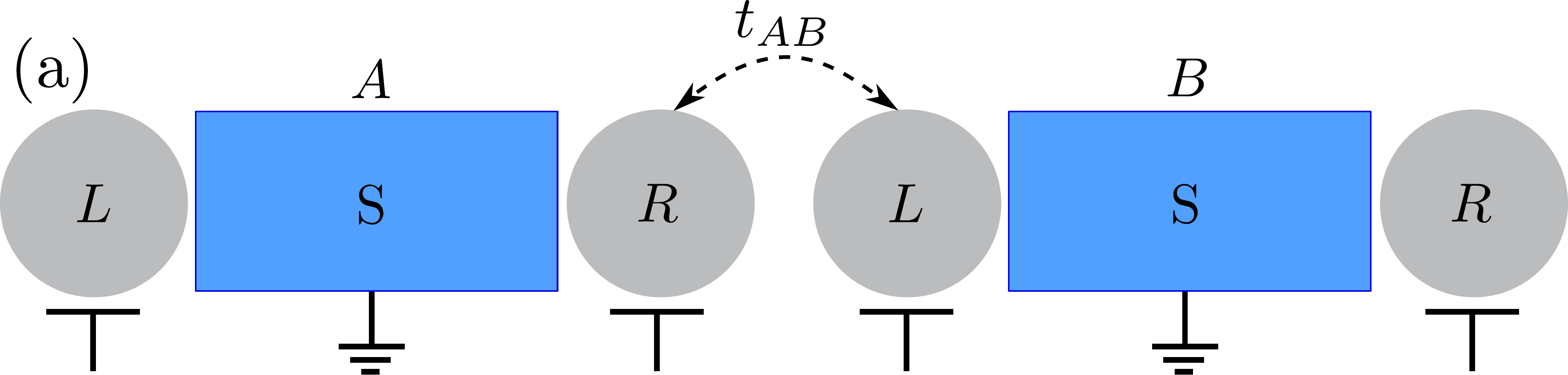}
\includegraphics[width=0.92\linewidth,trim={0cm -1cm 0cm 0cm},clip]{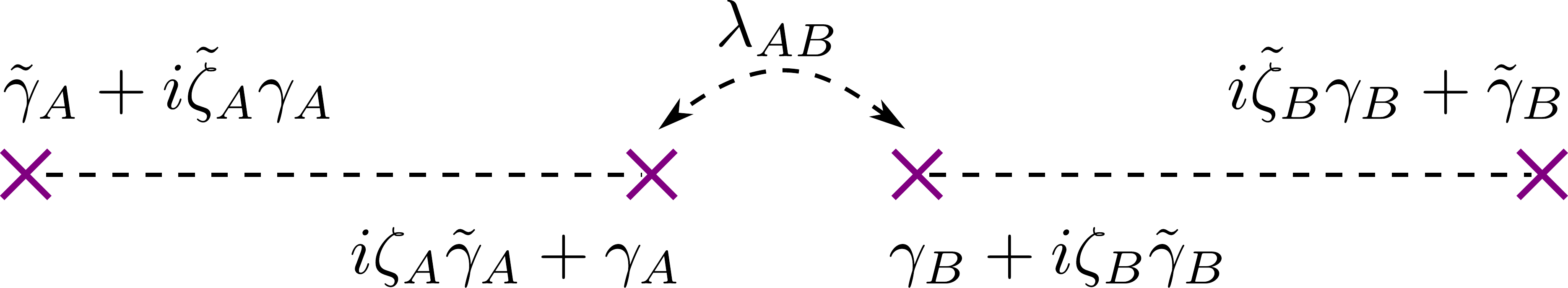}
\includegraphics[width=0.49\linewidth,trim={0cm -0.1cm 0cm 0cm},clip]{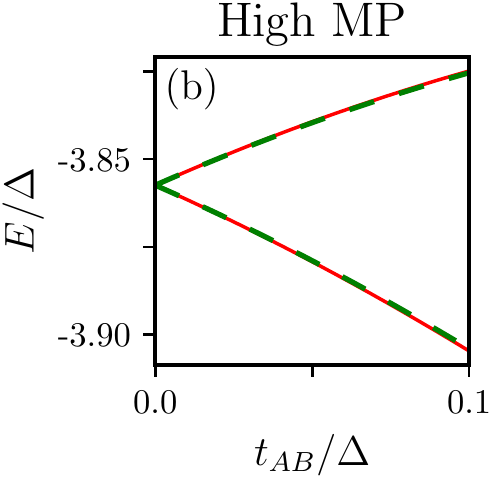}
\includegraphics[width=0.49\linewidth,trim={0cm -0.1cm 0cm 0cm},clip]{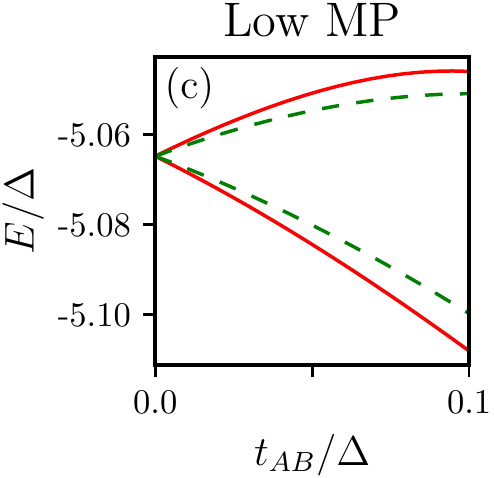}
\includegraphics[width=1\linewidth,trim={0cm -1.0cm 0cm 0cm},clip]{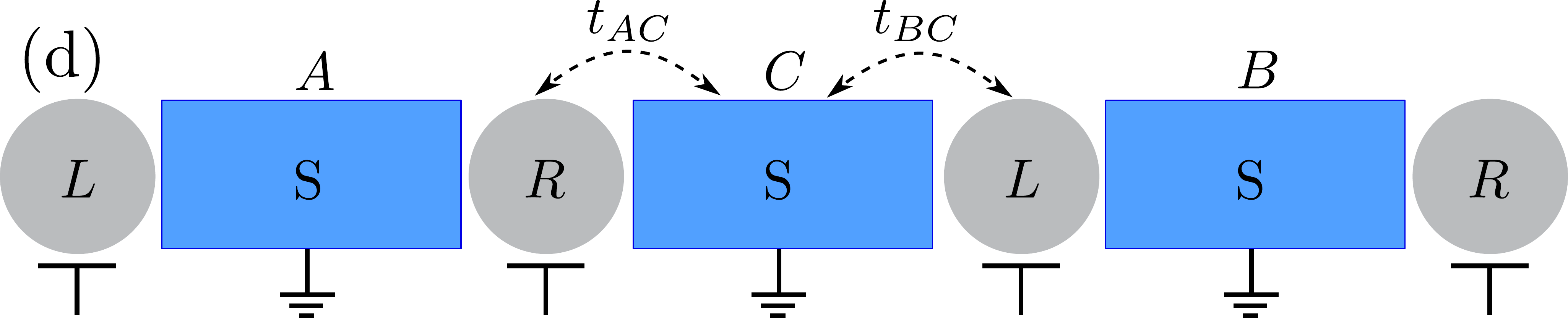}
\includegraphics[width=0.49\linewidth]{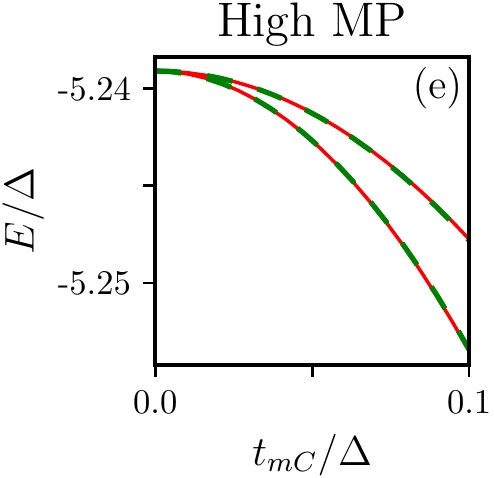}
\includegraphics[width=0.49\linewidth]{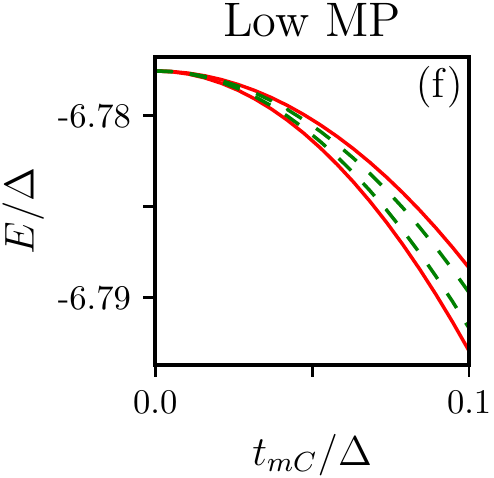}
\caption{(a) Sketch of two coupled PMM systems which can be used as a PMM qubit, with the two inner QDs coupled via direct tunneling (upper panel) and sketch of a low-energy (MBSs-only) model (lower panel). (b) and (c) The four lowest eigenenergies of the system in (a) plotted as a function of the QD coupling $t_{A B}$ at a high-MP sweetspot in (b) and a low-MP sweetspot in (c) (same parameters as the high-/low-MP sweetspots in Fig.~\ref{FigPC} and using $t_{AB} = 5 t_{AB}^\mathrm{SO}$). The full red (green dashed) lines show states with total even (odd) parity. (d) Same as (a), but with the inner QDs coupling via a superconducting segment hosting an ABS (and using $t_{AC} = 5 t_{AC}^\mathrm{SO}$). (e) and (f) Same as (b) and (c), but for the coupling mechanism in (d) ($m=A,B$).
}\label{fig:twoDQDs}
\end{figure}

When the coupling between the PMM systems is switched on, the four-fold degenerate ground state splits as seen in Figs.~\ref{fig:twoDQDs}(b, c) for the direct coupling [Eq.~(\ref{eq:HTAB})] and in Figs.~\ref{fig:twoDQDs}(e, f) for the coupling via an additional superconducting segment [Eq.~(\ref{eq:tildeHTAB})]. For small coupling strength (compared with the couplings within each PMM system), the main difference between the two coupling mechanisms is that for coupling via an additional superconducting segment, the effective coupling between QDs $AR$ and $BL$ is of second order in the couplings $t_{mC}$ and $t_{mC}^{{\rm SO}}$ to the ABS (although the details will depend on the energy of the ABS). We furthermore compare the cases of high and low MP. Figures~\ref{fig:twoDQDs}(b) and (e) show that for a system tuned to a high-MP sweet spot, switching on the coupling splits the four-fold degenerate ground state into two (nearly) two-fold degenerate states. This corresponds to an equal splitting within the even (red full lines) and odd (green dashed lines) parity subspaces. In contrast, a lower MP leads to a different splitting within the total even and total odd parity subspaces, and thus to a complete breaking of the ground-state degeneracy, see Figs.~\ref{fig:twoDQDs}(c) and (f). 

We can understand the MP-dependence of the spectra by comparing with the low-energy model introduced in Eq.~(\ref{eq:Hbar}) generalized to two coupled PMM systems
\begin{align}\label{eq:Hbar_ab}
\begin{split}
    H^\mathrm{lowE}_{AB} = {} & {}\frac{i}{2} \sum_{s=A,B} \xi_s \gamma_s \tilde{\gamma}_s \\
    + {} & {} \frac{i}{4} \left[ \lambda_{A B} (\gamma_A - i \zeta_A \tilde{\gamma}_A) (\gamma_B + i \zeta_B \tilde{\gamma}_B) - \hc \right],
\end{split}
\end{align}
where we have chosen the Majorana operators such that the ground states on the inner QDs ($AR$ and $BL$) are dominated by $\gamma_s$, but contain also an additional small fraction $\zeta_s$ of $\tilde{\gamma}_s$ (see Sect.~\ref{sec:lowE}). This model is illustrated in the lower panel of Fig.~\ref{fig:twoDQDs}(a). Formally, we can obtain the coupling term from the spinless model by projecting the electron operators $d_{AR}^{(\dagger)}$ and $d_{BL}^{(\dagger)}$ onto the low-energy fermion in Eq.~(\ref{eq:fminus}) and then re-expressing the result in terms of the Majorana operators in Eq.~(\ref{eq:Hbar}). 
The coupling $\lambda_{AB}$ can also induce terms $\propto \gamma_s \tilde{\gamma}_s$,
but they will scale as $\lambda_{A B}^2$ and we will neglect them in the following where we focus on small $\lambda_{A B}$. The connection between $\zeta$ and the MP is given by Eq.~(\ref{eq:lowEMP}), although this expression should not be used for a very small Zeeman energy (leading to a small MP) where the low-energy model is not appropriate. 




As mentioned above, the coupling between the PMMs depends on the phase of $\lambda_{A B}$, which can be controlled if the superconductors are connected in a loop. Just as in Fig.~\ref{fig:twoDQDs}, we focus below on real $\lambda_{A B} > 0$, where  Eq.~(\ref{eq:Hbar_ab}) simplifies to 
\begin{align}\label{eq:Hbar_ab_real}
\begin{split}
    H^\mathrm{lowE}_{AB} = {} & {}\frac{i}{2} \sum_{s=A,B} \xi_s \gamma_s \tilde{\gamma}_s \\ 
    + {} & {} \frac{i \lambda_{A B}}{2} (\gamma_A \gamma_B + \zeta_A \zeta_B \tilde{\gamma}_A \tilde{\gamma}_B). 
\end{split}
\end{align}

If each PMM system is at a sweet spot when the coupling is switched off, then $\xi_s = 0$ by definition. If we in addition have perfect PMMs at the sweet spot, the MP is unity and $\zeta_s = 0$, and Eq.~(\ref{eq:Hbar_ab_real}) shows that turning on $\lambda_{A B}$ leads to an equal splitting within the subspaces with total even and total odd parity, see Figs.~\ref{fig:twoDQDs}(b,e). This splitting corresponds to a finite energy of the fermion $f_\alpha = \frac{1}{2} \gamma_A + \frac{i}{2} \gamma_B$, but a preserved two-fold degeneracy associated with the uncoupled fermion $f_\beta = \frac{1}{2} \tilde{\gamma}_A - \frac{i}{2} \tilde{\gamma_B}$ (see Appendix \ref{app:fusion_outcome} for an explanation for the specific forms of $f_\alpha$ and $f_\beta$). For imperfect PMMs---MP less than unity and $|\zeta_s| > 0$---also $\tilde{\gamma}_{A}$ and $\tilde{\gamma}_B$ couple, leading to a finite energy of the fermion $f_\beta$ and thus to a breaking of the remaining two-fold degeneracy. As a consequence, the splittings within the subspaces with even and odd total parity are no longer equal. This is what is seen in Figs.~\ref{fig:twoDQDs}(c,f) as a splitting between the full red and dashed green lines [note that for an even ground state one must take $\zeta_A = -\zeta_B$ in Eqs.~(\ref{eq:Hbar_ab_real})]. 

We note here that there is, in fact, an important advantage in using phase control to achieve a real $\lambda_{A B}$. Compared with Eq.~(\ref{eq:Hbar_ab}), the terms that are linear in $\zeta_s$ have vanished in Eq.~(\ref{eq:Hbar_ab_real}). Thus, a real $\lambda_{A B}$ reduces the effect of imperfect PMMs (low MP).

It is exactly this additional splitting for finite $\zeta_s$ that will be the deciding factor for whether the braiding protocols in Sect.~\ref{sec:braiding} work or not. It is therefore desirable to accurately measure the breaking of the ground-state degeneracy in two coupled PMM systems. One way would be to simply measure the spectrum for a finite coupling with transport spectroscopy (similar to Sect.~\ref{sec:conductance}). However, we will now show that by operating the coupled PMM systems as a qubit in both the total even parity and total odd parity subspaces, the breaking of the ground-state degeneracy can instead be measured in the time-domain, which should allow for detecting much smaller energy splittings. 

\subsection{Coherent control of Majorana qubits}\label{sec:qubits}
We start by considering a qubit encoded in the subspace with even total parity, such that $|ee\rangle$ is at the north pole and $|oo\rangle$ is at the south pole of the Bloch sphere. Then, detuning one or both PMM systems away from the sweet spot with the coupling between them switched off causes a rotation around the $z$-axis, while coupling the two PMM systems results in a rotation around an axis in the $xy$-plane. This becomes clear if we approximate the coupled PMM systems by the low-energy four-MBSs Hamiltonian~(\ref{eq:Hbar_ab_real}). If we choose 
\begin{align}
\begin{split}
    \gamma_{s} = f_s^\dagger + f_s, \; \tilde{\gamma}_s = i(f_s^\dagger - f_s),
\end{split}
\end{align}
with $f_s^\dagger$ being the creation operator for a fermion in the lowest-energy mode in PMM system $s=A,B$, then
$\ket{ee}$ is the ground state of the uncoupled system for $\xi_s > 0$, and we identify the Pauli matrices as $\sigma_z = -i \gamma_A \tilde{\gamma}_A = -i \gamma_B \tilde{\gamma}_B$, $\sigma_y = i \gamma_A \gamma_B = - i \tilde{\gamma}_A \tilde{\gamma}_B$, $\sigma_x = -i \gamma_A \tilde{\gamma}_B = -i \tilde{\gamma}_A \gamma_B$ (where the equality signs should be interpreted as the operators having the same effect within the total even parity subspace). Thus, within the total even parity subspace Eq.~(\ref{eq:Hbar_ab_real}) becomes 
\begin{align}\label{eq:Hbar_ab_even}
\begin{split}
H^\mathrm{lowE}_{AB,e} &= -\frac{\xi_+}{2} \sigma_z +  \frac{\lambda_{A B}}{2} (1 - \zeta_A \zeta_B)\sigma_y,
\end{split}
\end{align}
where $\xi_+ = \xi_A + \xi_B$. 

We now discuss how to initialize, control, and read out the PMM qubit. For simplicity, we base our discussion on the low-energy model in Eq.~(\ref{eq:Hbar_ab_even}), but the same operations work also when this is not a good approximation. The qubit can be initialized in an eigenstate of $\sigma_z$ by letting $\lambda_{A B} \rightarrow 0$ and detuning the QDs away from the sweet spot to make $\xi_A \neq 0$ and $\xi_B \neq 0$. This can be followed by either a parity measurement as described in Sect.~\ref{sec:initialization} or simply by letting the system relax (by quasiparticle poisoning) to the ground state, before tuning back to the sweet spot. Two-axis control of the qubit is achieved by pulsing $\xi_s$ (resulting in a rotation around the $z$-axis) or $\lambda_{A B}$ (resulting in a rotation around the $y$-axis). Readout is performed in the same way as initialization with $\lambda_{A B}$ switched off. 

Even if one has braiding as the end-goal, qubit measurement is an important stepping stone. In addition to being a test of initialization, coupling and readout, it allows measurements of the coherence time ($T_2$ time), which will be a limiting factor also for braiding. This can be done simply by measuring the decay of coherent oscillations with a finite $\lambda_{A B}$, or in a Ramsey-type experiment where one first performs a $\pi/2$ rotation into the $xy$-plane, then waits for some time before applying another $\pi/2$ rotation, giving a decaying signal as a function of the waiting time. This experiment in itself is not sensitive to $\zeta$ and we can equally well implement a qubit based on PMMs with low MP, although there is a slight lifting of the protection from variations in QD orbital energies which will likely affect the coherence times at very low MP. 

For the purpose of estimating $\zeta$ (or the MP) we now consider a qubit encoded instead in the subspace of total odd parity where we choose $\sigma_z = - i \gamma_A \tilde{\gamma}_A = i \gamma_B \tilde{\gamma}_B$, $\sigma_y = i \gamma_A \gamma_B = i \tilde{\gamma}_A \tilde{\gamma}_B$, $\sigma_x = i \gamma_A \tilde{\gamma}_B = -i \tilde{\gamma}_A \gamma_B$ (where the equality signs should be interpreted as the operators having the same effect within the total odd parity subspace). Then we obtain 
\begin{align}\label{eq:Hbar_ab_odd}
\begin{split}
    H^\mathrm{lowE}_{AB,o} &= -\frac{\xi_-}{2} \sigma_z +  \frac{\lambda_{A B}}{2} (1 + \zeta_A \zeta_B)\sigma_y,
\end{split}
\end{align}
where $\xi_- = \xi_A-\xi_B$. 

Comparing the second terms in Eqs.~(\ref{eq:Hbar_ab_even}) and~(\ref{eq:Hbar_ab_odd})  shows that for the same value of $\lambda_{A B}$, the effective coupling strength is different in the even and odd subspaces. This reflects the different splitting between the two green dashed lines compared with the splitting between the two red full lines in Figs.~\ref{fig:twoDQDs}(c,f). Thus, the difference in qubit rotation frequency around the $y$-axis for a given $\lambda_{A B}$ between the qubit in the even and odd subspaces provides a time-domain measurement of the Majorana quality.

It is important to note that at the sweet spot and with the coupling between the PMM systems switched off, the total even and total odd parity ground states are degenerate. Thus, the total even and total odd parity qubits will be operated with all parameters identical, the only difference during operation will be in the initialization. To initialize in the total odd subspace, we detune one PMM system away from the sweet spot to give an even-parity ground state (which can, for example, be done by setting $\varepsilon_{AL} = \varepsilon_{AR}$) but the other one to give an odd-parity ground state (which can, for example, be done by setting $\varepsilon_{BL} = -\varepsilon_{BR}$).

\subsection{Testing Majorana fusion rules}\label{sec:fusion}
A Majorana fusion experiment has been suggested as a probe of nonabelian physics that is experimentally easier than braiding~\cite{Aasen_PRX2016, Hell_milestonesPRB16, Clarke2017Apr, Beenakker2019, Zhou2020, Zhou2022, Souto2022, Liu2022_PMMfusion}. We here briefly discuss and comment on a fusion protocol for PMMs that is similar in spirit to that in Ref.~\cite{Aasen_PRX2016} and analogous to the recent proposal in Ref.~\cite{Liu2022_PMMfusion}. The aim of a fusion protocol is to initialize Majorana pairs and thereafter measure them in a different pairing configuration. The possible outcomes of this measurement (fusion) and the associated probabilities are a fundamental property of nonabelian anyons. The simplest version of the fusion protocol requires four PMMs and can be accomplished in either of the setups sketched in Figs.~\ref{fig:twoDQDs}(b,d) [where $U_{AB} \approx 0$ is needed for the setup in (b)] through the following steps: 

{\it (i)} The protocol starts with a large coupling between the two PMM systems. If the MP is high the ground state is almost two-fold degenerate [Figs.~\ref{fig:twoDQDs}(b,e)], but we can break that degeneracy by detuning the outermost QDs, $AL$ and $BR$. The system is then allowed to relax to the unique ground state before bringing the levels of QDs $AL$ and $BR$ back to their sweet-spot values. With true topological MBSs, which is  described by the low-energy model in Eq.~(\ref{eq:Hbar_ab}) with $\zeta_A = \zeta_B = 0$, this corresponds to initializing the system in a state where the innermost ($\gamma_A$ and $\gamma_B$) and outermost ($\tilde{\gamma}_A$ and $\tilde{\gamma}_B$) PMM pairs have a definite parity. A PMM system with $|M_{AR}| < 1$ and/or $|M_{BL}| < 1$ can be approximated by $|\zeta_A| > 0$ and/or $|\zeta_B| > 0$, in which case all four PMMs couple and the initial state deviates from the ideal one.

{\it (ii)} In the next step we reduce the coupling between the two PMM systems. This should be done adiabatically with respect to higher-energy states [not shown in Fig.~\ref{fig:twoDQDs} but corresponding to excitations within each PMM system of the type in Fig.~\ref{fig:init_readout}(a,c)]. At the end of this step, the ground state is (almost) two-fold degenerate independent of the MP, with a possible splitting set by a combination of residual coupling (if this cannot be switched off completely) and finite energy of each individual PMM system if it has not been tuned perfectly to the sweet spot [corresponding to finite $\xi_s$ in Eq.~(\ref{eq:Hbar_ab}) and~(\ref{eq:Hbar_ab_real})]. The operation of switching off the coupling should be fast compared to the inverse of that energy scale. 

{\it (iii)} Finally, the state of the PMM pairs within each PMM system is read out. The readout can be done as described in Sect.~\ref{sec:initialization} by detuning away from the sweet spot with the coupling between PMM systems switched off. In the ideal case, this corresponds to reading out the eigenvalue of $i \gamma_s \tilde{\gamma}_s$ in Eqs.~(\ref{eq:Hbar_ab}) and~(\ref{eq:Hbar_ab_real}). 

Thus, in the ideal case the fusion protocol will initialize the system in an eigenstate of $i \gamma_A \gamma_B$ and $i \tilde{\gamma}_A \tilde{\gamma}_B$, and then read out this state in the eigenbasis of $i \gamma_A \tilde{\gamma}_A$ and $i \gamma_B \tilde{\gamma}_B$. According to the Majorana fusion rules, the outcome of the readout is 50/50 (because of conservation of total parity, there are two possible outcomes). 

There is no principal problem associated with implementing this protocol in the setup with imperfect PMMs. The problem is instead that the protocol will work also for low-MP PMMs. In fact, as was noted in Ref.~\cite{Clarke2017Apr}, the outcome of a fusion protocol as sketched above (or any of the versions that are based on the same principle) will always be 50/50, as long as the protocol is diabatic with respect to the small ground-state splitting mentioned in step \textit{(ii)} above, but adiabatic with respect to all higher excited states. We show this in detail in Appendix~\ref{app:fusion_outcome}, where we demonstrate that the low-energy model in Eq.~(\ref{eq:Hbar_ab_real}) produces a 50/50 outcome independent of the values of $\zeta_s$, i.e., independent of the MPs. Thus, the protocol serves as a time-domain probe of the exactness of the ground-state degeneracy and is a useful test of adiabatic control of the PMM system, but it does not give conclusive evidence of the actual Majorana nature of the associated (near) zero-energy fermionic states. Therefore, we now turn to braiding experiments. 

\section{Nonabelian signatures and Majorana braiding}\label{sec:braiding}

Exchanging two MBSs leads to a non-trivial change in the system state, which serves as direct evidence of their nonabelian properties. However, physically exchanging MBSs is challenging and hence various protocols have been proposed which have the same effect on the states encoded by the MBSs, but which are hopefully easier to realize experimentally.  
In this section, we adapt three such proposals for the PMM platform and investigate the impact of the unavoidable imperfect nature of the PMMs. Throughout this section, we describe the PMM systems with appropriately extended versions of the low-energy model in Eqs.~(\ref{eq:Hbar_ab}) and~(\ref{eq:Hbar_ab_real}), which implicitly means that we assume that inter-QD Coulomb interactions are negligible or can be compensated for by gating.

\subsection{Non-abelian signatures in charge-transfer-based protocols}

\begin{figure}\centering
\includegraphics[width=1\linewidth]{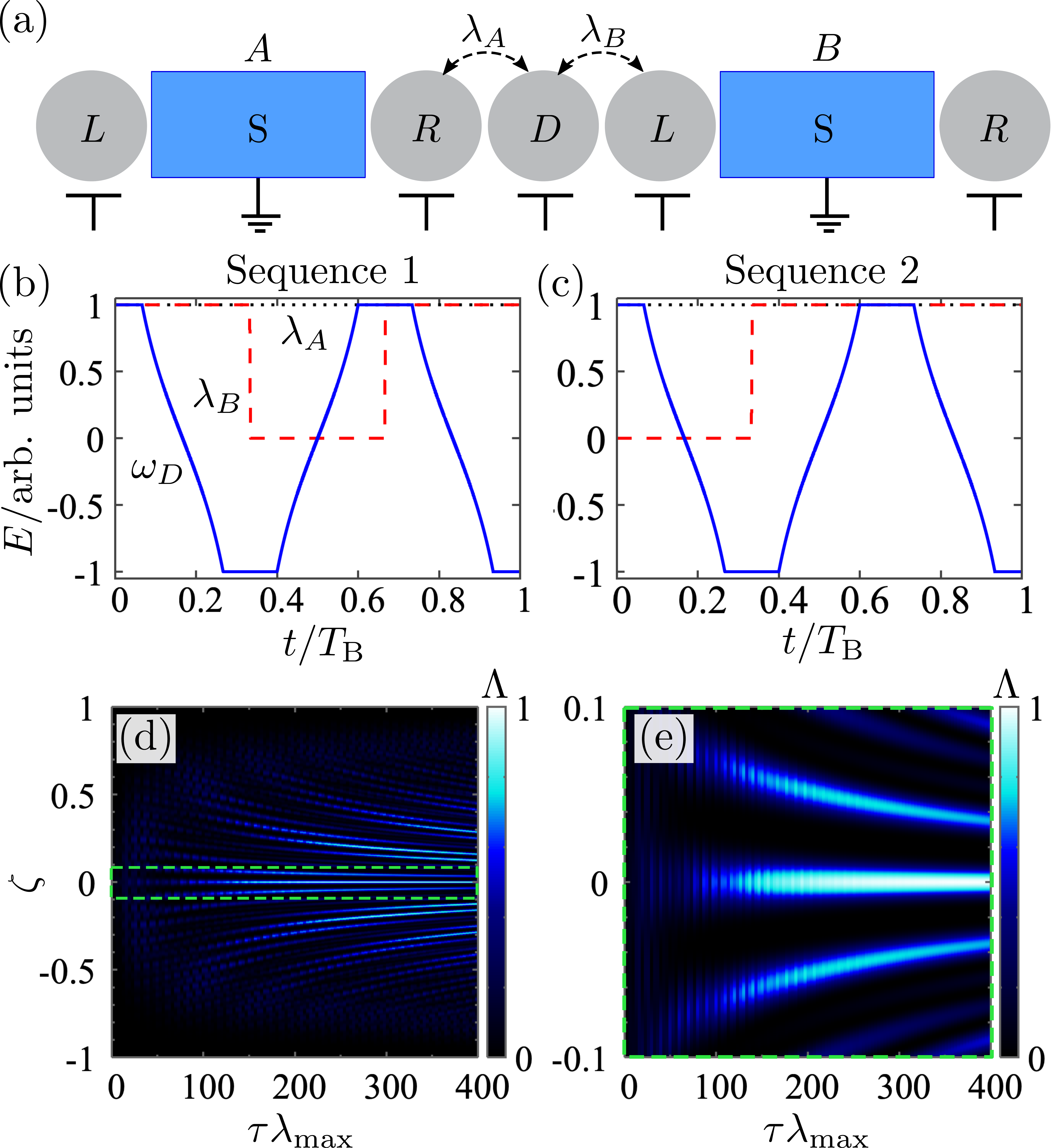}
\caption{Charge-transfer protocol for demonstration of PMM nonabelian properties. (a) Sketch of the setup with QD $D$ tunnel coupled to the two PMM systems. Panels (b) and (c) show the two sequences of the protocol, where the energy of QD $D$ is swept between positive and negative energies ($T_B$ is the duration of each sequence). To minimize the effect of dynamical phases, we sweep QD $D$ as $\omega_D(t)=\omega^{\rm max}_D\tan[\pi/4(t-t_0)/\tau]$, where $t_0$ is a time offset and $\tau$ controls the sweep time. Energies are normalized with respect to their maximum value, taken as $\lambda^{\rm max}_A=\lambda^{\rm max}_B=\omega^{\rm max}_D/500 = \lambda_\mathrm{max}$ (used as energy unit in the figure). (d) Protocol visibility $\Lambda$, defined as the product of the probabilities of finding the ideal Majorana outcome after sequences 1 and 2. For the initial state $|ee\rangle$,  $\Lambda=P^{1}_{oe} P^{2}_{eo}$. (e) Zoom-in of the visibility close to $\zeta=0$.}
\label{fig:charge-braiding}
\end{figure}

Charge-transfer based protocols~\cite{Flensberg_PRL2011, Souto2020,Krojer_PRB2022} provide conceptually simple tests of Majorana nonabelian properties. The basic setup (adapted to our PMM systems) is sketched in Fig.~\ref{fig:charge-braiding}(a) and the protocol is based on transferring charges between the central QD ($D$) and the PMM systems by sweeping the level of QD $D$ from negative to positive (or from positive to negative) energies. In an adiabatic operation, a single charge is transferred between QD $D$ and the coupled PMMs, thereby changing the joint fermion parity of the two PMM systems. When QD $D$ couples to a single PMM (for example $\gamma_{A}$), the operation can be understood mathematically as acting with the corresponding Majorana operator, $C_i=\gamma_{A}$ on the ground state of the PMM system. If QD $D$ is coupled to a pair of PMMs (say $\gamma_{A}$ and $\gamma_{B}$), the charge-transfer operation can be understood as acting with the operator $F_{AB}=(\gamma_A+\gamma_B)/\sqrt{2}$ on the PMM systems. Therefore, $B_{AB}=F_{AB}C_A$ gives the same result as braiding PMMs $\gamma_A$ and $\gamma_B$ and the difference between $C_AF_{AB}$ and $F_{AB}C_A$ is due to the nonabelian nature of the PMMs. 
It might be more convenient to add a third operation, {\i.e.} $F_{AB}C_AF_{AB}$ and $F_{AB}F_{AB}C_A$, so that, for perfect PMMs, each sequence gives a state with well defined PMM parities, differing only by the fermion between the left and right subsystems \cite{Krojer_PRB2022}.

To see how this protocol is affected by having PMMs with MP less than unity, we generalize the low-energy model in Eq.~(\ref{eq:Hbar_ab}) to include the extra QD $D$ in Fig.~\ref{fig:charge-braiding}(a):
\begin{align}
\begin{split}
    H=\omega_D d_D^\dagger d_D &+ \frac{i}{2} \sum_{s=A,B} \xi_s \gamma_s\tilde{\gamma}_s \\
    &+ \frac{i}{2}\sum_{s=A,B}\left[\lambda_{s}(\gamma_s -i \zeta_s \tilde{\gamma}_s)d - \mbox{H.c.}\right].\label{eq:Hchargetransfer}
\end{split}
\end{align}
The system is first initialized by detuning the PMM QDs, as explained in Sect. \ref{sec:initialization}. Sweeps of QD $D$ can flip the parity of the $A$ and $B$ PMM systems. To implement $F_{AB}$, we keep the tunnel rates to $A$ and $B$ equal ($\lambda_A=\lambda_B$), while $C_A$ can be achieved by setting $\lambda_B=0$. The two different sequences are sketched in Figs.~\ref{fig:charge-braiding}(b,c).

We test the protocol numerically by solving the time-dependent Schr\"odinger equation for the unitary time evolution while sweeping the level of QD $D$ up and down according to Figs.~\ref{fig:charge-braiding}(b,c). For these sweeps, we use a shape that minimizes the accumulation of dynamical phases \cite{Krojer_PRB2022}. For a system initialized in the state $|ee\rangle = |e\rangle_A |e\rangle_B$, the ideal outcome for sequence 1 (2) is $|oe\rangle$ ($|eo\rangle$), which can be measured using the parity readout described in Sect. \ref{sec:initialization}. Deviations from these outcomes are associated with imperfectness on the charge-transfer operations (as discussed in Refs.~\cite{Souto2020,Krojer_PRB2022}) and/or the less-than-unity MP (finite $\zeta_s$). Figure~\ref{fig:charge-braiding}(d) shows the protocol visibility, $\Lambda$, given by the product of the probabilities that the two sequences give the result expected for perfect Majoranas. 

The numerical calculations show small regions where $\Lambda\approx1$, including the PMM sweet spot, $\zeta=0$. The failure of the protocol for larger $\zeta$ is due to the finite splitting of the degenerate manifold which depends linearly on $\zeta$. This splitting introduces a dynamical phase and, therefore, a dependence of the final state on the details of how operations are performed. Therefore, a minimization of the dynamical phase would require fast operations, although slow enough to avoid non-adiabatic effects (which lead to failure of the protocol for too fast sweeps, even for $\zeta = 0$). This trade off results in a relatively narrow range close to $\zeta = 0$ where the designed sequences of operations can show Majorana nonabelian properties. In principle, it might be possible to extend the $\zeta$ range where the protocol is successful  by echoing away the dynamical phase, for example by reversing the sign of $\zeta$ between 
operations (similar to the echo used to compensate for an imperfectly fine tuned phase in Ref.~\cite{Krojer_PRB2022}).

\subsection{Measurement-based braiding}

\begin{figure} \centering
\includegraphics[width=1\linewidth, trim={0.0cm 1.cm 0.0cm 0.0cm},clip]{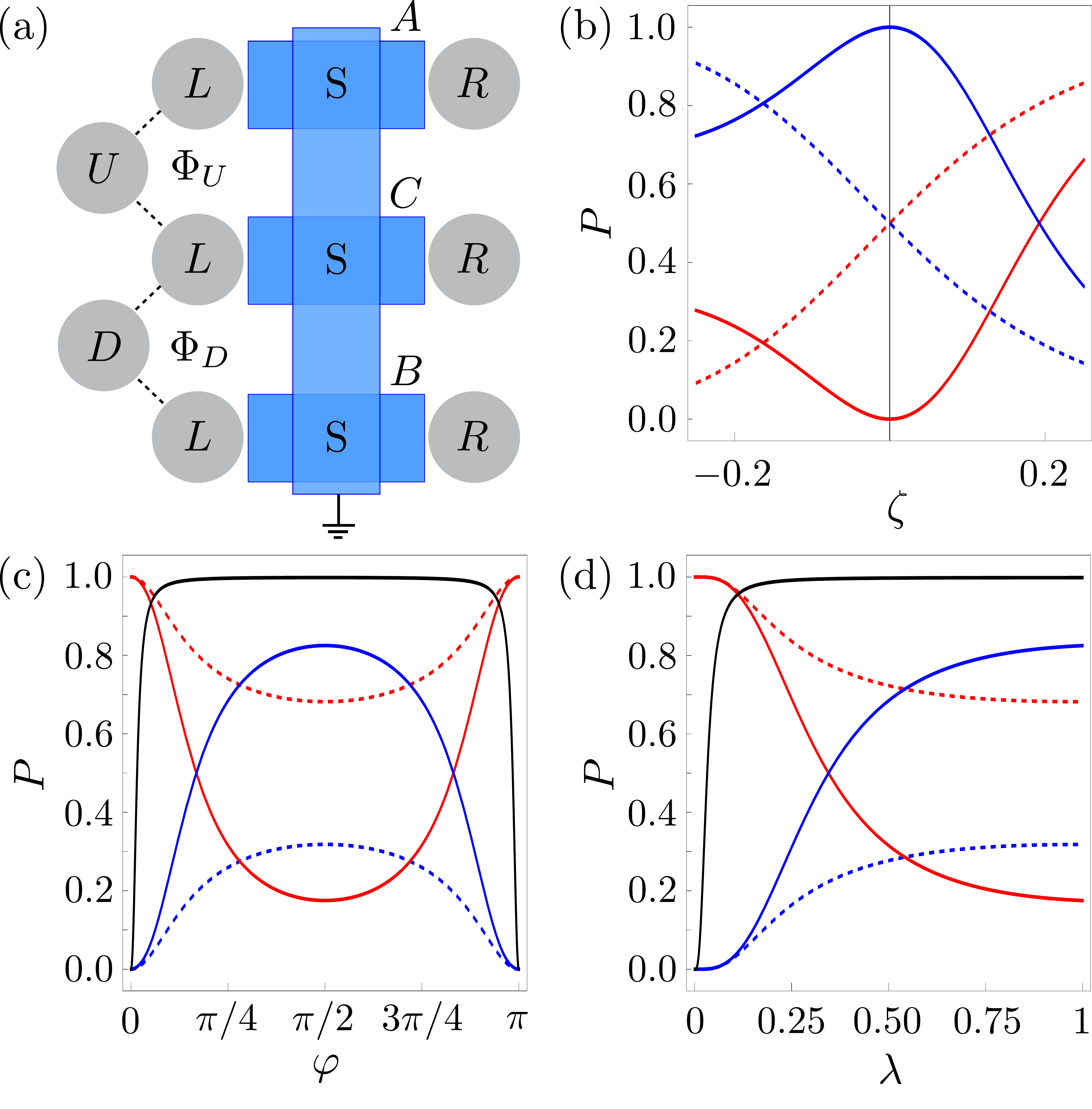}
	\caption{\label{fig:measbraid} Measurement-based braiding. (a) Sketch of the proposed device consisting of three PMM systems ($A,B,C$) and two readout QDs ($U,D$). (b)--(d) Outcomes of the braiding protocols for single (dashed line) and double (solid lines) braids, starting from the state $|eee\rangle = |e\rangle_A |e\rangle_C |e\rangle_B$. The lines represent the weights ($P$) of the final states. The red curves give the probability for staying in the initial state, $|eee\rangle$, while the blue curves give the probability for ending up in the other even state, $|oeo\rangle$. (b) Braiding outcome as a function of $\zeta = \zeta_A = \zeta_B = \zeta_C$ with $\lambda=1$ and $\phi=\pi/2$. (c) Braiding outcome as a function of $\phi$ for $\lambda=1$ and $\zeta=0.1$. (d) Braiding outcome as a function of $\lambda$ with $\varphi=\pi/2$ and  $\zeta=0.1$. The full black lines show for comparison the probability of the double braid to end up in the $|oeo\rangle$ state with very high-quality PMMs ($\zeta=0.01$). For this close-to-ideal case, the results are nearly independent on $\lambda$ and $\phi$.}
\end{figure}

Another way of effectively performing braiding and operations on a set of MBSs is by using a measurement-based protocol \cite{Bonderson_PRL2008, Plugge_NJP2017, Karzig_PRB2017}. A protocol that uses QDs to read out the parity of pairs of MBSs and performs a braiding was presented in Ref.~\cite{Karzig_PRB2017}. Here, we discuss how this protocol can be adapted to PMMs. The setup shown in Fig.~\ref{fig:measbraid}(a) contains three PMM systems ($A,B,C$) and two additional QDs for parity readout ($U,D$). PMM system $s$ hosts two PMMs, $\gamma_s$ localized mainly on QD $sL$ which is connected to one or two readout QDs, and $\tilde{\gamma}_s$ which is localized mainly on QD $sR$ and is not connected to any readout QD. PMM systems $A$ and $B$ together have a doubly degenerate ground state for a given parity which can be manipulated by braiding. PMM system $C$ provides a pair of auxiliary PMMs.

Let us start with describing the procedure for the case where all PMMs have unit MP. The aim will be to braid $\gamma_A$ and $\gamma_B$. This can be done through the following sequence of measurements: \textit{(i)} initialize the joint parity of $\gamma_C$ and $\tilde{\gamma}_C$ to be even; \textit{(ii)} measure the joint parity of $\gamma_A$ and $\gamma_C$; \textit{(iii)} measure the joint parity of $\gamma_B$ and $\gamma_C$; \textit{(iv)} measure the joint parity of the auxiliary PMMs $\gamma_C$ and $\tilde{\gamma}_C$ which we postselect for an even outcome. After this sequence, one has performed an operation identical to a braiding of $\gamma_A$ and $\gamma_B$ if the outcomes of measurements \textit{(ii)} and \textit{(iii)} are both even, see Ref.~\cite{Karzig_PRB2017} (for different outcomes of \textit{(ii)} and/or \textit{(iii)}, the result deviates from the ideal braiding by a phase). More formally, the operator $B_{AB}$ for braiding $\gamma_A$ and $\gamma_B$ can be realized by the following set of projections:
\begin{equation}\label{B12def}
	B_{AB} \propto M_{C \tilde{C}} M_{BC} M_{AC} M_{C \tilde{C}},
\end{equation}
where $M_{s s'}$ ($M_{s \tilde{s}}$) denotes a projective measurement onto the state with even parity of $\gamma_s$ and $\gamma_{s'}$ ($\gamma_s$ and $\tilde{\gamma}_s$). 

The projection $M_{C \tilde{C}} = (1+i\gamma_C\tilde{\gamma}_C)/2$ is done as described in Sect.~\ref{sec:initialization} and can in principle be perfect independent of the MP. We now discuss readout of the joint parity of $\gamma_A$ and $\gamma_C$ using readout QD $U$, which is done along the lines described in Refs. \cite{Flensberg_PRL2011, Plugge_NJP2017, Karzig_PRB2017, munk2020parity, steiner2020readout}. The readout of the parity of $\gamma_B$ and $\gamma_C$ is completely analogous. For MP less than unity, these readouts will not be perfect, which will lead to deviations from the ideal braiding result.

The readout QD $U$ is coupled to QD $AL$ ($CL$) with tunnel amplitude $\lambda_{AU}$ ($\lambda_{CU}$) and the coupled system can be described by a Hamiltonian completely analogous to Eq.~(\ref{eq:Hchargetransfer}).
First we note that the coupling of the PMM systems to a single-level QD can only decrease the degeneracy by 2. This is because a single fermion coupled to a degenerate subspace consisting of $N$ fermions will necessarily yield $N-1$ dark fermions~\cite{Geier2023}. The readout QD $U$ thus splits the system according to the occupation of the fermionic mode which it couples to 
%
%
and after some rather lengthy algebra (see Appendix \ref{app:meas}), one can express the parity operator which couples to QD $U$ in terms of the parameters $\zeta_s$, $\phi = \arg(\lambda_{AU}/\lambda_{CU})$, and $\lambda = |\lambda_{AU}/\lambda_{BU}|$. When the coupling is switched on, it will constitute a measurement of this parity operator. If the coupling is not too strong, the system ends up in one of its two energy eigenstates \cite{munk2020parity, steiner2020readout,Szechenyi_PRB2020, Schulenborg_PRB2023}. The maximal visibility turns out to be achieved for $\phi = \pi/2$ and $\lambda = 1$.

This projection varies between the ideal topological case discussed above for all $\zeta_s=0$ to the ``trivial'' (or fermionic) case for $\zeta_s=1$. 
However, the fermionic case is actually not as trivial as one might expect. For $\zeta_s=1$ and the initial state $|oee\rangle = |o\rangle_A |e\rangle_C |e\rangle_B$ the braiding protocol \eqref{B12def} gives $B_{AB}|oee\rangle=|eeo\rangle$, maybe as expected. However, a double braid (i.e., acting twice with $B_{12}$) annihilates the state. The braiding protocol also annihilates $ |eee\rangle$ and $|eeo\rangle$, whereas $|oeo\rangle$ is an eigenstate of $B_{12}$. To conclude, the braiding operation does not necessarily give the ``trivial'' braiding of electrons in the fermionic limit. 

The protocol gives the outcome expected for topological MBSs only for all $\zeta_s=0$. In the small-$\zeta$ regime, there will some deviations from the ideal braiding results. This is shown in Fig.~\ref{fig:measbraid}(b) where the outcomes of single and double braids are shown as a function of $\zeta = \zeta_A = \zeta_B = \zeta_C$. Figures~\ref{fig:measbraid}(c,d) show the stability of the braiding outcome when changing $\phi$ and $\lambda$. For ideal PMMs, the outcome of the protocol does not depend on these parameters (although the visibility of the readout signal does, such that readout might become difficult far from $\phi = \pi/2$ and $\lambda = 1$).

To conclude this part, the presented measurement-based protocol could in principle be used to demonstrate PMM braiding. However, the interpretation might become difficult because the results will depend strongly on the device parameters, such as magnetic fluxes, gate-voltage settings and tunnel couplings. On the other hand, this dependence might be a very useful way to characterize the device.

\subsection{Hybridization-induced braiding}

\begin{figure*}\centering
\includegraphics[width=0.9\linewidth]{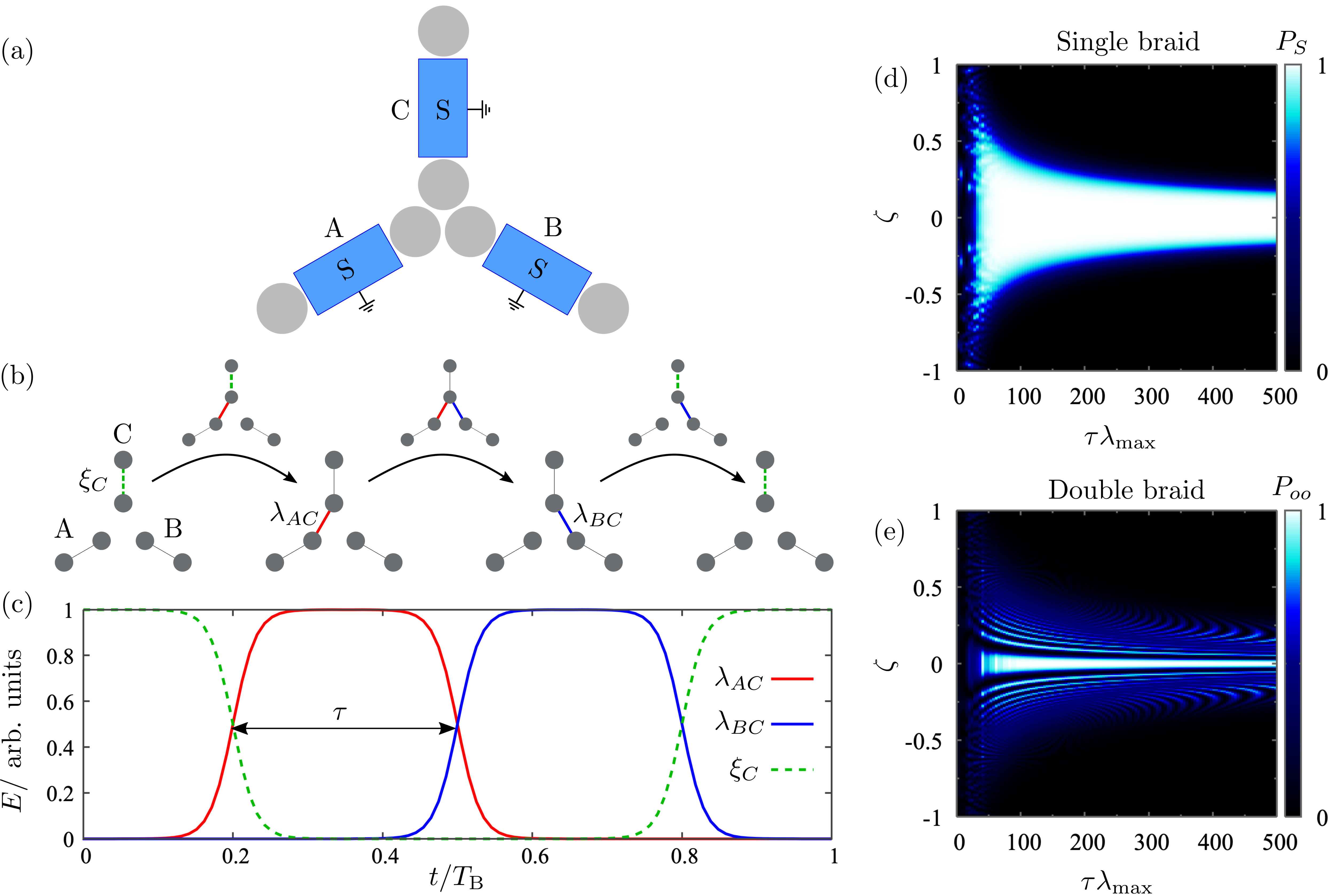}
\caption{Hybridization-induced braiding. (a)~Sketch of setup. (b)~Braiding protocol, where the coupling between PMM pairs are switched on and off. (c)~Strengths of the different couplings during the protocol, as a function of time $t_\mathrm{B}$ normalized by the protocol duration $T_\mathrm{B}$. (d)~$P_S$ [as defined in Eq.~\eqref{eq:P_1braid}] after a single braid, as a function of  $\zeta$, and the time of each interaction pulse $\tau$ (in units of the inverse of $\lambda_\mathrm{max}$, the maximal value of $\xi_C$, $\lambda_{AC}, \lambda_{BC}$). (e)~Probability of measuring a parity flip of the PMM systems $A$ and $B$ ($= P_{oo}$ with the chosen initial conditions) after a double braid. We used max$(\lambda_{AC}) = $max$(\lambda_{BC})=$max$(\xi_C)$, which are switched on and off using a sigmoid function with a rate $r=50/\tau$.}
\label{fig:T-braiding}
\end{figure*}

The hybridization-induced braiding protocol relies on alternating couplings between MBSs to effectively exchange their positions. There are many versions of this protocol in the literature for braiding of topological MBSs~\cite{Clarke2011, van_Heck_NJP2012, Karzig2015, Aasen_PRX2016, Hell_milestonesPRB16, Hell2017, Clarke2017Apr}
, which differ from each other mainly in the physical mechanism used to realize MBS coupling, initialization and readout. Our proposed PMM version is based on the setup in Fig.~\ref{fig:T-braiding}(a) with three PMM systems and resembles the setup in Ref.~\cite{Boross_2023} (which, however, focused on real-space braiding). To describe this system, we use a generalization of Eq.~(\ref{eq:Hbar_ab_real})
\begin{equation}\label{eq:H_Tbraiding}
    H=\sum_{s}\frac{i\,\xi_s}{2}\gamma_s\tilde{\gamma}_s+\frac{i}{2}\sum_{ss'}\lambda_{ss'}(\gamma_s \gamma_{s'} + \zeta_s \zeta_{s'} \tilde{\gamma}_s \tilde{\gamma}_{s'})\,,
\end{equation}
where $ss'=AB,BC,AC$ and we have taken the coupling terms $\lambda_{ss'}$ to be real, which requires phase control of the superconductors. As in Sect.~\ref{sec:2DQD}, $\zeta_s$ is a measurement of the relative amplitude of the outer-QD PMMs $\tilde{\gamma}_s$ on the inner coupled QDs, such that $\zeta_s = 0$ for unit MP.

The hybridization-induced braiding protocol requires tuning the couplings between three PMMs~\cite{Clarke2011, van_Heck_NJP2012, Karzig2015, Aasen_PRX2016, Hell_milestonesPRB16, Hell2017, Clarke2017Apr}, and the different steps involved are schematically illustrated in Fig.~\ref{fig:T-braiding}(b). In the first step, all couplings between PMM systems are switched off (all $\lambda_{s s'} = 0$) and the system is initialized in a state with a well-defined fermion parity in subsystems $A$ and $B$ which are tuned to the Majorana sweet spot ($\xi_A = \xi_B = 0$). The ground state degeneracy of the central subsystem is split $\xi_C \neq 0$ by coupling the two PMMs $\gamma_C$ and $\tilde{\gamma}_C$, which is represented by a green dashed line in Fig.~\ref{fig:T-braiding}(b). The initialization and degeneracy breaking within each PMM system is described in Sect.~\ref{sec:initialization}. 

The aim of the protocol is then to exchange PMMs $\gamma_A$ and $\gamma_B$. During the protocol, $\lambda_{AC}, \lambda_{BC}$, and $\xi_C$ are switched on and off in a specific sequence, as shown in Figs.~\ref{fig:T-braiding}(b,c). At each step one or two (but never three or zero) couplings are switched on. For PMMs with unit MP ($\zeta_A = \zeta_B = \zeta_C = 0$) this guarantees that the ground state remains two-fold degenerate within a subspace of fixed total parity. The entire protocol is carried out one time (single braid) or two times (double braid), after which the parity of subsystem $A$ and/or $B$ is read out (see Sect.~\ref{sec:initialization}) with all $\lambda_{ss'} = 0$.

For definiteness we assume that the system is initialized in the state $|ee\rangle = |e\rangle_A |e\rangle_B$ (in the calculation also PMM system $C$ is initialized to be even, but this is not important and we suppress it in our notation). The ideal result of exchanging PMMs $\gamma_A$ and $\gamma_B$ once (one sequence of the braiding protocol) is to transform the state according to 
\begin{equation}\label{eq:single_braid}
|ee\rangle \rightarrow \frac{1}{\sqrt{2}}\left( |ee\rangle + i|oo\rangle \right).
\end{equation}
Carrying out the same sequence twice brings $\gamma_A$ and $\gamma_B$ back to their original positions, but the state is transformed according to
\begin{equation}\label{eq:double_braid}
|ee\rangle \rightarrow  |oo\rangle.
\end{equation}

Readout of the parity of either PMM system $A$ or $B$ is sufficient to detect successful braiding, but readout of both provides a consistency check, and readout of also PMM system $C$ can be used to verify that total parity has remained fixed during the protocol (no quasiparticle poisoning). We imagine that the experiment is carried out many times and compare the outcome with the ideal probability distributions from Eqs.~(\ref{eq:single_braid}) and~(\ref{eq:double_braid}). For the single braid, we define the function
\begin{equation}
    P_S=16P_{oo}\left(1-P_{ee}\right)P_{ee}\left(1-P_{oo}\right)\,.
    \label{eq:P_1braid}
\end{equation}
where $P_{ee}$ ($P_{oo}$) is the probability to measure even (odd) parity in PMM systems $A$ and $B$. $P_S = 1$ for the ideal 50/50 outcome and $P_S = 0$ for the trivial outcome of both PMM systems always being even or always being odd. For the double braid we simply have  $P_{oo} = 1$ in the ideal case, while $P_{oo} = 0$ in the trivial case. 

We simulate the braiding protocol by solving the time-dependent Schr\"odinger equation for the model in Eq.~(\ref{eq:H_Tbraiding}) with the initial conditions described above. Figure~\ref{fig:T-braiding}(c) shows the calculated result for $P_S$ defined in Eq.~(\ref{eq:P_1braid}) after a single braid, as a function of $\zeta = \zeta_A = \zeta_B = \zeta_C$ and $\tau$, the duration of each interaction. Figure~\ref{fig:T-braiding}(c) shows the analogous result for $P_{oo}$ after a double braid.

For $\zeta=0$, we find the result expected for topological MBSs, $P_S = 1$ and $P_{oo} = 1$ for the single and double braid respectively. The result of the single braid seems relatively stable for $\zeta \lesssim 0.5$. This is because the quantities that are read out ($P_{ee}$ and $P_{oo}$) are insensitive to the phase between the $|ee\rangle$ and $|oo\rangle$ components of the wavefunction in Eq.~(\ref{eq:single_braid})~\cite{Clarke2017Apr}. However, this phase manifests itself after a second braid operation, and the double braid result is therefore much more sensitive to $\zeta$. But also for the double braid, there are special values of the operation speed where a close to ideal result is found even for rather large $\zeta$. Thus, we conclude that to prove nonabelian exchange with PMMs, one should do both single and double braid protocols, and also vary the speed of the protocol to verify the stability of the result. 


\section{Conclusions}
In this work, we have presented a roadmap for next-generation experiments on PMMs in minimal Kitaev chains. We believe that the ultimate goal of such experiments should be the demonstration of the nonabelian nature of Majoranas. However, the roadmap contains a few milestones along the way, including assessment of PMM quality (closeness to true topological Majoranas),  initialization and readout of fermion parity encoded in PMMs (including measurements of quasiparticle poisoning times), and coherent control of a PMM qubit and measurements of coherence times.  

We have presented three different braiding-like tests of Majorana nonabelian physics, focusing on how the PMM quality affects the outcome. The suggestion for charge-transfer-based nonabelian operations is simplest in terms of the required setup, which can be realized in a linear geometry. However, it has the disadvantage of only being partially protected (even for topological Majoranas) and we also show that it only produces a nontrivial result if the PMMs have very high quality. The protocol for measurement-based braiding becomes equivalent to topologically protected braiding for perfect PMM quality, but it can produce nonabelian (but unprotected) results also for low-quality PMMs, and care is needed to avoid false positive outcomes of an experiment. The hybridization-induced braiding most closely resembles the picture of braiding as moving Majoranas around each other, is fully protected for topological Majoranas and gives a nontrivial outcome for lower-quality PMMs compared with the charge-transfer protocol, but requires a comparatively complex geometry and operational protocol. 

For all protocols, we emphasize that a true signature of nonabelian physics must exhibit some kind of stability to be distinguishable from conventional manipulation of a quantum state. For the charge-transfer and hybridization-based protocols, this stability manifests in the result being independent of protocol speed (within some interval). For the measurement-based protocol, the stability is instead with respect to parameters characterizing the coupling between the PMMs and the readout device. 

With the Majorana delocalization parameter $\zeta = 0.1$ the charge-transfer protocol fails, the measurement-based braiding gives results that deviate by $\sim 10-20 \%$ from the ideal result and have some stability to variations in phase and tunnel coupling, while the hybridization-induced braiding approaches the ideal stable results even for a double braid (with a single braid being successful also for much larger $\zeta$). Based on Eq.~(\ref{eq:lowEMP}), $\zeta = 0.1$ corresponds to an MP of around 0.98, which seems well within experimental reach (the high-MP case in Figs.~\ref{FigPC}--\ref{fig:twoDQDs} has MP of around 0.986). With $\zeta = 0.01$, all three protocols give a stable and close-to-ideal result. However, this corresponds to an MP of around $0.9998$, which in our spinful model would require $E_Z \approx 20 \Delta$ if the rest of the parameters are chosen as in Figs.~\ref{FigPC}--\ref{fig:twoDQDs} (but significantly higher charging energies would reduce this value~\cite{Tsintzis2022}). For all protocols, nonabelian results without significant stability appear already for much larger $\zeta$, which may not be unambiguous evidence for topologically protected nonabelian physics of Majoranas, but exciting new physics nonetheless. 

It would be an interesting direction for future theoretical works to try to optimize the braiding protocols presented here to improve their stability and extend their usefulness to PMM systems with lower MP. This could, for example, involve optimized pulses and/or echo pulses to minimize or cancel dynamical phases. It would also be interesting to theoretically investigate decoherence in PMM systems and its effect on different braiding protocols. Another relevant question is how the proposed experiments would be affected by using longer QD chains (which are already being pursued experimentally~\cite{Bordin2023}).

In conclusion, although there are clearly challenges involved in experimental tests of nonabelian physics with PMMs, we believe that this work has shown promising paths towards this goal.

\section{Acknowledgments}
This work has received funding from the European Research Council (ERC) under the European Unions Horizon 2020 research and innovation programme under Grant Agreement No. 856526, the Spanish CM “Talento Program” (project No. 2022-T1/IND-24070), the Swedish Research Council under Grant Agreement No. 2020-03412, the European Union’s Horizon 2020 research and innovation program under the Marie Sklodowska-Curie Grant Agreement No. 10103324, NanoLund, and the Novo Nordisk Foundation, Grant number NNF22SA0081175, NNF Quantum Computing Programme. The computations were enabled by resources provided by the National Academic Infrastructure for Supercomputing in Sweden (NAISS) at PDC, the Center for High-Performance Computing at the Royal Institute of Technology (KTH), partially funded by the Swedish Research Council through Grant Agreement No. 2022-06725.

A.T. and R.S.S. contributed equally to this work.

\bibliography{bibliography}

\appendix

\section{Fusion protocol outcome}\label{app:fusion_outcome}
In this Appendix, we show that the outcome of the  fusion protocol suggested in Section \ref{sec:fusion} is $50/50$ regardless of the weight $\zeta_{A,B}$ of the outer MBSs $\tilde{\gamma}_{A,B}$---mainly localized on $AL$ and $BR$---on the inner QDs $AR$ and $BL$ (cf.~Fig.~\ref{fig:twoDQDs}). For definiteness, we consider the even parity subspace similarly to what we did in the main text.  For an operation that is diabatic with respect to the ground-state splitting but adiabatic with respect to excited states, the probabilities for the fusion outcomes are given by the matrix elements:
\begin{equation}\label{eq:fusion_prop}
    P_{n_An_B} = |\braket{e_\alpha e_\beta}{n_A n_B}|^2,
\end{equation}
\noindent where $\ket{e_\alpha e_\beta}$ is taken to be the ground state for the two coupled PMM systems modeled by the Hamiltonian given in Eq.~(\ref{eq:Hbar_ab_even}) and $\ket{n_A n_B}$ correspond to the (almost) degenerate ground state for the uncoupled systems, right after step \textit{(ii)} of the fusion protocol ($n_A,  n_B = e_A$ or $o_A$). $\ket{e_\alpha e_\beta}$ is annihilated by the fermions:
\begin{equation}
    f_{\alpha} = \frac{1}{2} \gamma_A + \frac{i}{2} \gamma_B, \quad
    f_{\beta} = \frac{1}{2} \tilde{\gamma}_A - \frac{i}{2} \tilde{\gamma}_B,
\end{equation}
\noindent which can be easily checked using the final result of this section. $\ket{e_A e_B}$ is annihilated by the fermions:
\begin{equation}
    f_{A} = \frac{1}{2} \gamma_A + \frac{i}{2} \tilde{\gamma}_A , \quad
    f_{B} = \frac{1}{2} \gamma_B  + \frac{i}{2} \tilde{\gamma}_B.
\end{equation}
\noindent We can read off the probabilities in Eq.\ (\ref{eq:fusion_prop}) by writing the ground state of the coupled systems in the basis $\ket{n_A n_B}$. The eignevalues of Eq. (\ref{eq:Hbar_ab_even}) at the sweet spot ($\xi_+ = 0$) are
\begin{equation}
    v_\pm = \pm \frac{\lambda_{A B}}{2}(1 - \zeta_A \zeta_B),
\end{equation}
with the corresponding normalized eigenvectors
\begin{equation}
    V_{\pm} = \frac{1}{\sqrt{2}} 
    \begin{pmatrix}
    1  \\
    \pm i  
    \end{pmatrix}.
\end{equation}
The ground state can be written as
\begin{equation}
    \ket{e_\alpha e_\beta} = \frac{1}{\sqrt{2}} [\ket{e_A e_B} - i \ket{o_A o_B} ],
\end{equation}
and $P_{ee} = P_{oo} = \frac{1}{2}$ for any $\zeta_{A,B}$. The fusion outcome is thus independent of the values of $\zeta_{A,B}$ and thus the MP of the PMM systems. Although Eq. (\ref{eq:Hbar_ab_even}) is derived for a real and positive $\lambda_{AB}$, the conclusion is more general and is also valid for a complex $\lambda_{AB}$.

\begin{widetext}
\section{Quantum dot readout for the measurement-based braiding protocol}
\label{app:meas}

In this Appendix, we give the details on how to calculate the projective measurements done by the readout QD $U$ for the measurement-based braiding. The Hamiltonian is 
analogous to Eq.~(\ref{eq:Hchargetransfer}) and the QD $U$ electron is described by the operator $d_U =(\gamma_U+i\tilde{\gamma}_U)/2$. Writing this in first-quantized form in the Majorana basis  $\{\gamma_U,\tilde{\gamma}_U,\gamma_A,\tilde{\gamma}_A,{\gamma}_C,\tilde{\gamma}_C\}$ the Hamiltonian becomes (up to a constant)
\begin{equation}\label{HL15Majbasis}
	H = \frac{i|\lambda_{AU}|}{2}
	\left(
	\begin{array}{cccccc}
		 0& \bar{\omega}_U& 0& \zeta_A &\lambda \sin(\phi)& \zeta_C\lambda \cos(\phi)  \\
		 -\bar{\omega}_U& 0& -1&0& -\lambda \cos(\phi)& \zeta_C\lambda \sin(\phi)  \\
		 0& 1& 0& 0& 0& 0  \\
		 -\zeta_A&0& 0& 0& 0& 0 \\
		 - \lambda\sin(\phi)& \lambda \cos(\phi)& 0& 0& 0& 0  \\
		-\zeta_C\lambda \cos(\phi) & -\zeta_C\lambda \sin(\phi) & 0& 0& 0& 0  	
	\end{array}\right),
\end{equation}
where $\bar{\omega}_U=\omega_U/|\lambda_{AU}|$ with $\omega_U$ the orbital energy of QD $U$, $\lambda=|\lambda_{CU}/\lambda_{AU}|$, and $\phi=\arg(\lambda_{CU}/\lambda_{AU})$. The two linear combinations of Majorana modes that couple to QD $U$ are:
\begin{equation}\label{gammalc}
	\begin{aligned}
		a &= \zeta_A\tilde{\gamma}_A+ \lambda\sin(\phi){\gamma}_C+ \zeta_C\lambda\cos(\phi)\tilde{\gamma}_C,\\
		b &= {\gamma}_A+\lambda\cos(\phi){\gamma}_C- \zeta_C\lambda\sin(\phi)\tilde{\gamma}_C.
	\end{aligned}
\end{equation} 
After orthonormalization of these two vectors, we have that the fermion, whose parity is read out by QD $U$, is in Majorana representation given by  $f_U^\mathrm{read}=(\gamma_U^\mathrm{read}+i\tilde{\gamma}_U^\mathrm{read})/2$ where 
\begin{equation}\label{gamma1s}
	\begin{aligned}
		\gamma_U^\mathrm{read} = & (a-x  b)\frac{\zeta_C^2 \lambda ^2 \sin^2(\phi )+\lambda ^2 \cos ^2(\phi )+1}{\left(\zeta_A^2+\lambda ^2 \sin ^2(\phi )\right) \left(\zeta_C^2 \lambda ^2 \sin ^2(\phi )+1\right)+\lambda ^2 \cos ^2(\phi ) \left(\zeta_A^2+2 \zeta_C^2 \lambda ^2 \sin ^2(\phi )+\zeta_C^2\right)+\zeta_C^2 \lambda ^4 \cos ^4(\phi )},\\
		\tilde{\gamma}_U^\mathrm{read} &= \frac{b}{\sqrt{1+\lambda^2(\zeta^2_C\sin^2(\phi)+\cos^2(\phi))}}
		,\\
		x = & \frac{\lambda^2\sin(\phi)\cos(\phi)(1-\zeta_C^2)}{1+\lambda^2(\zeta^2_C\sin^2(\phi)+\cos^2(\phi))}
	\end{aligned}
\end{equation} 
The parity being projected by the measurement is thus
\begin{equation}\label{pgeneral}
	p_U= i\gamma_U^\mathrm{read}\tilde{\gamma}_U^\mathrm{read}.
\end{equation}
We note that for the case of perfect PMMs, $\zeta_A = \zeta_C=0$, one gets 
$\gamma_U^\mathrm{read}\propto -\lambda\sin(\phi)\cos(\phi)\gamma_A+\sin(\phi)\gamma_C$ and $\tilde{\gamma}_U^\mathrm{read}\propto \gamma_A+\lambda \cos(\phi) \gamma_C$. This means that $i\gamma_U^\mathrm{read} \tilde{\gamma}_U^\mathrm{read} \propto i \gamma_A\gamma_C$ as expected, independent of $\lambda$ and $\phi$.

When performing the calculations of the results of the braiding protocol, we define the fermion operators, $d_s=(\gamma_s+i\tilde{\gamma}_s)/2$ with $s=A,B,C$. These operators then act on the Fock basis states $\ket{n_A n_C n_B}$, which allows us to write the matrices for the various projectors.
\end{widetext}
\end{document}